\documentclass[11pt]{article}
\usepackage{pdfpages}
\pdfoutput=1

\usepackage{amssymb}

\usepackage{amsmath}
\usepackage{amsfonts}
\usepackage{enumerate}
\usepackage{amsthm}

\newcommand{\bB}{\mbox{\boldmath $B$}}

\newcommand{\bX}{\mbox{\boldmath $X$}}

\newcommand{\bE}{\mbox{\boldmath $E$}}

\newcommand{\bQ}{\mbox{\boldmath $Q$}}
\newcommand{\bzero}{\boldmath 0}

\newcommand{\bbeta}{\mbox{\boldmath $\beta$}}

\newcommand{\bPsi}{\mbox{\boldmath $\Psi$}}

\begin{document}

\title{{\Large  Quantile Functional Regression using Quantlets}
\author{Hojin Yang, Veerabhadran Baladandayuthapani,  Jeffrey S. Morris \\
Department of Biostatistics\\ 
 The University of Texas MD Anderson Cancer Center
 }
 } 
\date{Oct 31, 2017} 
\maketitle

\mbox{}
\vspace*{-0.4 in}



\begin{center}
\textbf{Abstract}
\end{center}
In this paper, we develop a quantile functional regression modeling framework that models the
distribution of a set of 
common repeated observations
from a subject through the quantile function, which is regressed on a set of covariates
to determine how these factors affect various aspects of the underlying subject-specific distribution.
To account for smoothness in the quantile functions, we introduce custom basis functions we call \textit{quantlets} that are sparse, regularized, near-lossless, 
and empirically defined, adapting to the features of a given data set and containing a Gaussian subspace so {non-Gaussianness} can be assessed.   
While these quantlets could be used within various functional regression frameworks, we build a Bayesian framework
that uses nonlinear shrinkage of quantlet coefficients to regularize the functional regression coefficients and allows fully Bayesian inferences after fitting a Markov chain Monte Carlo.  Specifically, we apply global tests to assess which covariates have any effect on the distribution at all, followed by local tests to identify at which specific quantiles the differences lie while adjusting for multiple testing, and to assess whether the covariate affects certain major aspects of the distribution, including location, scale, skewness, Gaussianness, or tails.  If the difference lies in these commonly-used summaries, our approach can still detect them, but our systematic modeling strategy can also detect effects on other aspects of the distribution that might be missed if one restricted attention to pre-chosen summaries.   We demonstrate the benefit of the basis space modeling through simulation studies, and illustrate the method using a biomedical imaging data set in which we relate the distribution of pixel intensities from a tumor image to various demographic, clinical, and genetic characteristics.

\vspace*{.3in}

\newpage

\section{ {\bf Introduction} }

With the advent of automated measurement devices, it is increasingly common to observe a large number of repeated measurements for each subject or other experimental unit in a study.  Some examples include cancer imaging data consisting of intensity measurements for a large number of pixels,
activity monitoring data consisting of a large number of activity level measurements over time, and climate data consisting of various climate variables such as temperature and rainfall measured daily over a long period of time.  While it is sometimes of interest to model the spatial or temporal relationships among these measurements, at other times one is interested in looking at the subject-specific distribution of these measurements and associating a set of covariates with aspects of these distributions, which include among others the mean, median, variance, skewness, heavy-tailedness, and various upper and lower quantiles. 
In these cases, the most common analytical approach is to compute these pre-defined summaries and perform separate regression analyses for each.

\begin{figure}[!htb]
\centering
\caption{\small Differences in density distributions: 
panel A reveals four densities
black Normal($\mu=1, \sigma=5$), blue Normal($\mu=1, \sigma=10$),
green Normal($\mu=10, \sigma=10$), and red  Skewed Normal ($\mu=10, \sigma=10$)  and 
their corresponding cumulative density functions and quantile functions are shown in panels B and C, respectively. 
  \label{S0_Figure}}
\includegraphics[height=2.2in,width=5.6in]{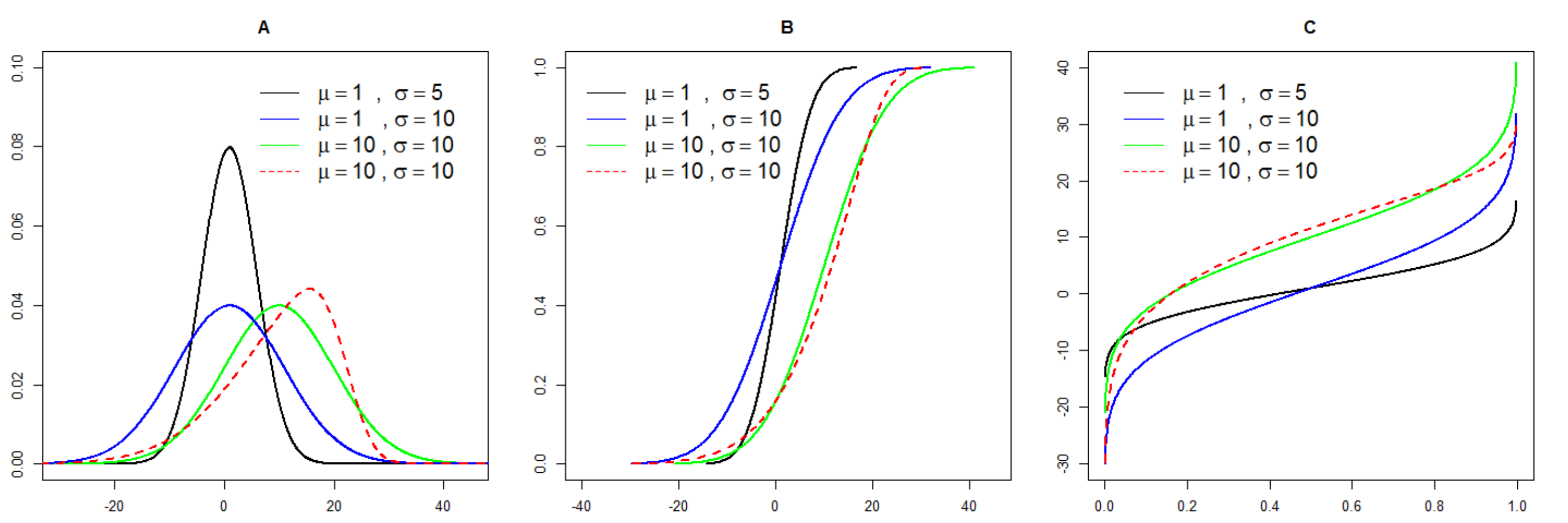}
\vspace{0.5cm} 
\end{figure}

While this strategy is reasonable and often can yield meaningful results, it has numerous drawbacks.   The exploratory regression analysis of numerous different summaries raises multiple testing questions, and if the key distributional differences are not contained in the pre-defined summaries, then this approach can miss out on important insights.  To illustrate this point, consider the densities plotted in panel (a) of Figure \ref{S0_Figure}. 
We see that just extracting the mean from the entire density function cannot distinguish two distributions for which  
the means are identical but one is more variable than the other (black solid line and blue solid  line).
Similarly,  note the red dashed line and green solid lines mark densities with identical means and variances, so only considering these 
summaries would miss out on their difference in skewness.  Also, solely looking at the center of a distribution via the mean or median can miss out
on differences in the tails, which in some settings can be the most scientifically relevant parts of the distributions.
It would be preferable to model the entire distribution, thus retaining the information in the data and potentially finding any differences.

There are various choices for representing the subject-specific distributions, including the distribution function, the quantile function, or the density function (if it exists).  The three panels of Figure \ref{S0_Figure} show all three for the example distributions.  In this paper, we choose to represent and model the distribution through the quantile function, which has numerous advantages as described in Section 2.6, including a fixed, common domain $[0,1]$, their ease of estimation by order statistics without any need for a smoothing parameter, and the ability to readily compute distributional moments.  Thus, our approach is to represent each subject's data via their empirical quantile function $Q_i(p), p\in {\cal P}= [0,1]$, computed from the order statistics, and then treat these as functional responses regressed on a set of scalar covariates $x_{ia}; j=1,\ldots,A$ through $Q_i(p) = B_0(p) + \sum_{a=1}^J x_{ia} \beta_a(p) + E_i(p)$.  This models the \textit{distribution of subject-level distributions} as a function of subject-level covariates.  We call the fitting of this model \textit{quantile functional regression}, which is clearly
different and distinguished
  from other forms of quantile regression in existing literature in Section 2.1.

One simple approach to fitting this model would be to interpolate each subject's data onto a common grid of ${\cal P}$ and then perform independent regressions for each interior point $p$.  This would lead to estimators that are unbiased but inefficient, as they would not borrow strength across nearby $p$, which should be similar to each other.  We refer to this strategy as \textit{naive quantile functional regression}.   As is typically done in other functional regression settings (see review article by \cite{morris2015functionalreg}), alternatively one could borrow strength across $p$ using basis representations, with common choices including splines, principal components, and wavelets.  In this paper, we will introduce a new strategy for construction of a custom basis set 
we call 
 \textit{quantlets} that is sparse, regularized, near-lossless, 
and empirically defined, adapting to the features of the given data set and containing the Gaussian distribution as a prespecified subspace so non-Gaussianness can be assessed.  Representing the quantile functions with a quantlet basis expansion, we propose a Bayesian modeling approach for fitting the quantile functional regression model that utilizes shrinkage priors on the quantlet coefficients to induce regularization of the regression coefficients $\beta_a(p)$, and leading to a series of global and local inferential procedures that can first determine whether $\beta_a(p)\equiv 0$ and then assess which $p$ and/or distributional summaries (e.g. mean/variance/skewness/Gaussianness) characterize any such difference.  While based on quantile functions, our model will also be able to provide predicted distribution functions and densities for any set of covariates to use as summaries for users more accustomed to interpreting densities than quantile functions.

 \begin{figure}[!htb]
 \centering
 \caption{\small Characterizing tumor heterogeneity from distributional summaries of MRI pixel intensities: 
the two graphs include kernel density estimates and the raw empirical quantile functions as representations of tumor heterogeneity (pixel intensities within the tumor); black line: female patient without DDIT3 mutation; red line: male patient without DDIT3 mutation; blue line: female patient with DDIT3 mutation; and green line: male patient with DDIT3 mutation. The images in other columns represent the T1-post contrast MRIs of the brains, with tumor boundaries indicated by black lines. 
  \label{S5_Figure_Intro}}
\includegraphics[height=4.2in,width=5.6in]{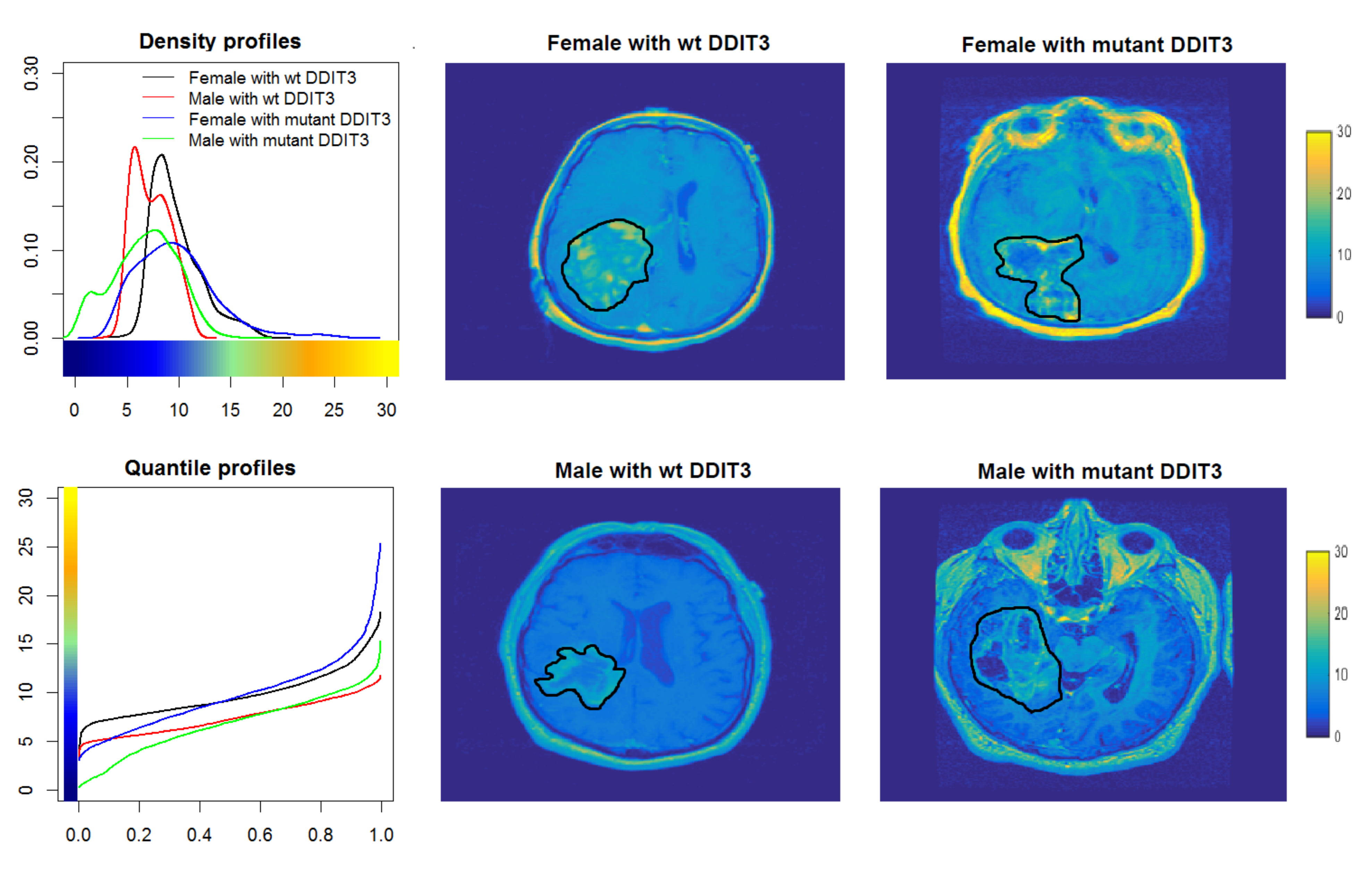}
\end{figure}

While broadly applicable, our methods are motivated by and illustrated using a Glioblastoma multiforme (GBM) data set 
in which we seek to relate aspects of the distributional characteristics of pixel intensities from patients' brain tumor images to clinical, demographic, and genetic factors.  The rightmost four plots of Figure \ref{S5_Figure_Intro} display MRI images for 4 patients with GBM, two males and two females, and with and without mutations in the DDIT3 gene, an important gene associated with GBM development, with tumor boundaries indicated by the black lines.  The bottom left plot contains the raw empirical quantile functions of the pixel intensities  within each of these four tumors and the upper left plot contains corresponding smoothed density estimates.  Features of these images may comprise clinically useful biomarkers, and one of the key types of features commonly investigated in cancer imaging include so-called \textit{histogram features}, which are summaries of these pixel intensity distributions \cite{just2014improving}.  It is of scientific interest to study the pixel intensity distributions for a set of 64 GBM tumors of which these four are a subset and investigate their associations with various covariates including age, sex, tumor subtype, DDIT3 mutation status, EGFR mutation status, and survival status ($>12$ months, $\le 12$ months).  Rather than extracting several distributional summaries and performing separate regressions, as is typically done, we aim to model the entire subject-specific pixel intensity distribution using our quantile regression framework after which we can succinctly characterize any observed differences.

We construct a set of {\it quantlet} basis functions for these data and use them to build a Bayesian quantile functional regression model including these covariates. This framework allows us to $(1)$ regress the quantile functions on the set of covariates, $(2)$ perform global tests to assess which covariates have an effect on the outcome distribution,
$(3)$ perform  local tests to identify which quantiles and/or features of the distribution are characterizing these differences, including various tests for moments of the outcome distribution on the covariates such as  mean, variance, and skewness.
Although in this article we utilize our novel framework to study GBM,  the framework can be applied to any setting in which one wishes to regress entire distributions on a set of covariates.

This paper is organized as follows. In Section 2, we introduce the general quantile function regression model, introduce \textit{quantlets}, describe how to construct a set of quantlet basis functions for a given data set, and describe our Bayesian approach to fitting the model.  In Section 3, we describe simulation studies conducted to evaluate the finite-sample performance of our method and demonstrate the benefit of incorporating quantlet bases in the modeling. In Section 4, we apply our method to real GBM data and perform various tests to obtain scientific results. We provide concluding remarks in Section 5.

\section{{\bf Models and Methods}}

\subsection{{\bf Quantile Functions and Empirical Quantile Functions}}  \label{sec:QF}

Let $Y$ be a real valued random variable and $F_{Y}(y)$ be its 
cumulative distribution function (right-continuous) such that $F_{Y}(y)=\text{P}(Y\le y)$,
and $p=F_{Y}(y)$ be the percentage of the population less than or equal to $y$. 
The quantile function of $Y$, defined for $p \in [0,1]$,  is defined as
 \begin{align*}
Q(p)=Q_{Y}(p)=F_{Y}^{-1}(p)=\inf{(y: F_{Y}(y) \ge p     )}.
 \end{align*}
Distributional moments are easily computable as simple functions of quantile function, for example with mean, variance, skewness, and kurtosis given by 
 \begin{align}  \label{Quant_formula} 
&\mu_{Y}  =  \text{E}(Y)= 
\int_{0}^{1} Q_{Y}(p)  dp     \notag \\ 
&{\sigma}_{Y}^2 =  \text{Var}(Y)=  
 \int_{0}^{1}(Q_{Y}(p) -\text{E}(Y) )^2 dp,   \notag \\ 
 &{\xi}_{Y}=  \text{Skew}(Y)= 
\int_{0}^{1}(Q_{Y}(p) -\text{E}(Y))^3/\text{Var}(Y)^{3/2} dp, ~~\text{and}     \notag  \\ 
&{\varphi}_{Y}=    \text{Kurt}(Y)= 
\int_{0}^{1}(Q_{Y}(p) -\text{E}(Y))^4/\text{Var}(Y)^2 dp.   
\end{align}
Let $Y_{(1)}\le \cdots\le Y_{(m)}$ be the order statistics 
 For $p \in [1/(m+1),m/(m+1)]$, the empirical quantile function of $Y$ is given by
  \begin{align*}
\widehat{Q}(p) =  (1-w)Y_{([(m+1)p])} + wY_{([(m+1)p]+1)},
 \end{align*}
 where 
 $[x]$ is an integer less than or equal to $x$
 and  $w$ is a weight such that $(m+1)p=[(m+1)p]+w$.
This empirical quantile function is an estimate of the true quantile function.

 As shown in \cite{parzen2004quantile}, for a fixed $p$, the empirical estimator  
is consistent and is asymptotically equivalent to a Brownian bridge 
when the density function $f_{Y}(y)$ exists and is positive.  This provides us an easy unbiased estimate for the subject-specific quantile function.
For this reason, we choose to represent each subject's data by their empirical quantile function and to study how the subject-specific distributions vary with covariates.  
In this paper, we are interested in studying outcomes $Y$ that are absolutely continuous, meaning that the corresponding quantile functions are continuous and smooth, without jumps that would occur for discretely valued random variables.
 For brevity, we omit the estimator notation for the empirical quantile functions and just refer to them as 
 $Q(p)$. 

\subsection{ {\bf Quantile Functional Regression Model}} \label{liter}

Suppose that for a series of subjects $i=1,\ldots,n$ we observe a sample of $m_i$ observations from which we construct a subject-specific empirical quantile function $Q_i(p_j)$ for $p_j=j/(m_i+1); j=1,\ldots, m_i$, along with a set of $A$ covariates  $\bX_i =(x_{i1},\dots, x_{iA})^{T}$.  
Note that by construction all subject-specific empirical quantile functions $Q_i(p_j)$ are non-decreasing in $p$.  See Section 4 of the supplement for further discussion of monotonicity issues in this framework.

The quantile functional regression model is given by
 \begin{equation}
Q_i(p)= \sum_{a=1}^{A}x_{ia}\beta_{a}(p) +  E_i(p)= \bX_i^{T}\bB(p) +  E_i(p),   
 \label{p5_qfm_i}  
\end{equation}
where  $\bB(p)=( \beta_1(p), \dots, \beta_A(p) )^{T}$ is a column vector of length $A$ 
containing unknown fixed functional coefficients for the quantile $p$ and 
   $E_i(p)$ is a residual error process, assumed to follow a mean zero Gaussian process with   
  the covariance surface, $\Sigma(p_1, p_2)=\text{cov}\{ E_i(p_1), E_i(p_2) \}$.  The structure of
  $\Sigma(\cdot,\cdot)$ captures the variability across subject-specific quantiles, and the diagonals capture
  the intrasubject covariance across $p$.   
In practice, we will focus our modeling on $p \in \mathcal{P} = [\delta, 1- \delta]$, with  $\delta=\max_{i\le n}\{1/(m_i+1)\}$ 
being the most extreme quantile estimable from the subject with the fewest observed data points.  In this paper,
we are primarily interested in settings with at least moderately large numbers of observations per subject, i.e. $m_i$ not too small, 
and in later studies will extend our work to sparse data settings with few observations per subject.

\begin{table}
\caption{Types of regression based on response type and objective function. 
      \label{objective}}
\resizebox{1.0\columnwidth}{!}{    
\begin{tabular}{   c| c c } \hline
& $\text{{\bf  Objective function}}$   &  $\text{{\bf  Objective function}}$ 
  \\ 
    $\text{{\bf Response}}$ $(\cdot)$
& $E( (\cdot) |X)$ &  $F_{(\cdot)}^{-1}(p|X)$         \\    \hline \hline
$\text{{\bf scalar}}$ $Y$	&$\text{{\bf classic regression}}$ 	
&$\text{{\bf quantile regression}}$	  \\
$\text{{\bf function}}$ $Y(t)$	&$\text{{\bf functional  regression}}$ 	
&$\text{{\bf functional quantile regression}}$	   \\
$\text{{\bf quantile function}}$ $Q(p)$
&$\text{{\bf quantile functional regression}}^{\ast}$	 	
&$\text{{\bf quantile functional quantile regression}}$ \\
	\hline	
		\hline	
\end{tabular}
}
\end{table}      

To place our model in the proper context within the current literature on quantile and functional regression, Table \ref{objective} lists various types of regression in terms of response and objective function. 
 In contrast to classical regression, which specifies the mean of the response conditional on a set of covariates, \textit{quantile regression} \cite{he2000quantile,koenker2005quantile,yang2015quantile} 
 works by estimating a pre-specified $p$-quantile of the response distribution conditional on the covariates, either with independent  \cite{koenker2004quantile,hao2007quantile,davino2013quantile} 
  or spatially/temporally correlated errors \cite{koenker2004quantile,reich2012bayesian,reich2012spatiotemporal}.  Most existing methods fit independent quantile regressions for each desired $p$, which can lead
  to crossing quantile planes, although recent methods (e.g. \cite{yang2017joint}) jointly model all quantiles, borrowing strength across $p$ using Gaussian process priors.  Parallel to these efforts are methods to perform \textit{Bayesian density regression} \cite{dunson2007bayesian}, in which the density of the response variable is modeled as a function of covariates via dependent Dirichlet processes  \cite{muller1996bayesian,maceachern1999dependent,griffin2006order,dunson2006bayesian}.
   These quantile regression models are inherently different from the setting of this paper, as they are modeling the quantile of the \textit{population} given covariates, while our framework is modeling the quantile function of each \textit{subject} as a function of subject-specific covariates.  Another difference is that, in general, these methods do not model intrasubject correlation in settings for which there is more than one observation per subject.

Other regression methods have been designed for functional responses.  There is a subset of the functional regression literature (see \cite{morris2015functionalreg} for an overview)
 that involve regression of a functional response on a set of covariates, with classical functional regression focusing on the mean function conditional on covariates 
 \cite{faraway1997regression,wu2000kernel,guo2002functional,ramsay2006functional,morris2006wavelet,reiss2010fast,goldsmith2011functional,goldsmith2011penalized,scheipl2015functional,meyer2015bayesian}, 
 and \textit{functional quantile regression} that computes
 the quantile of functional response conditional on covariates, using
  the check function as the objective function \cite{brockhaus2015functional,brockhaus2015fdboost}.  Again these methods are not modeling subject-specific, but rather population-level quantiles.
Other recent works on functional quantile regression have focused on the quantile of the scalar response distribution regressed on a set of functional covariates  
 \cite{ferraty2005conditional,cardot2005quantile,chen2012conditional,kato2012estimation,kato2012asymptotics,li2016inference}. 

All of these methods differ, fundamentally, from the quantile functional regression framework described in this paper.  For these methods, the quantile regression is computing the $p^{th}$ quantile of the population given covariates X, while in our case, we are interested in modeling the $p^{th}$ quantile of an individual subject’s distribution given X.  In our case, we are modeling the empirical quantile function for each subject as the response, and using a classical (mean) regression of these subject-specific quantile functions onto a set of scalar covariates, i.e. estimating the expected quantile function for a subject given a set of covariates. Note that it would also be possible to compute the $p^{th}$ quantile of the distribution
of specific empirical quantile functions for each p conditional on covariates, which could be dubbed \textit{quantile functional quantile regression}, but this model is not addressed in the current paper.

\subsection{{\bf Quantlet Basis Functions}} \label{sec:quantlets}

If all empirical quantile functions are sampled on (or interpolated onto) the same grid (i.e. $m_i \equiv m \forall i=1,\ldots,n$), then a simple way to fit model (\ref{p5_qfm_i}) would be to fit separate linear regressions for each $p$.  However, this naive approach would treat observations across $p$ as independent.  This leads to a regression model that fails to borrow strength across $p$, and thus is expected to be inefficient for estimation of the functional coefficients $\beta_a(p)$, and ignores correlation across $p$ in the residual error functions $E_i(p)$, which would adversely affect any subsequent inference.  We call this approach \textit{naive quantile functional regression} in our comparisons below.

Basis function representations can be used to induce smoothness across $p$ in $\beta_a(p)$ and capture intra-subject correlation in the residual error functions $E_i(p)$.  In existing functional regression literature, common choices for basis functions include splines, Fourier, wavelets, and principal components, and smoothness is induced across $p$ by regularization of the basis coefficients via L1 or L2 penalization \cite{morris2015functionalreg}.  Here, we introduce a strategy to construct a custom basis set called \textit{quantlets} for use in the quantile functional regression model that have many desirable properties, including regularity, sparsity, near-losslessness, interpretability, and empirical determination allowing them to capture the salient features of the empirical quantile functions for a given data set.

\begin{figure}[!htb]
\centering
\caption{Graphical illustration of the entire procedure for constructing the quantlets.
  \label{diagram}}
\includegraphics[height=2.5in,width=4.5in]{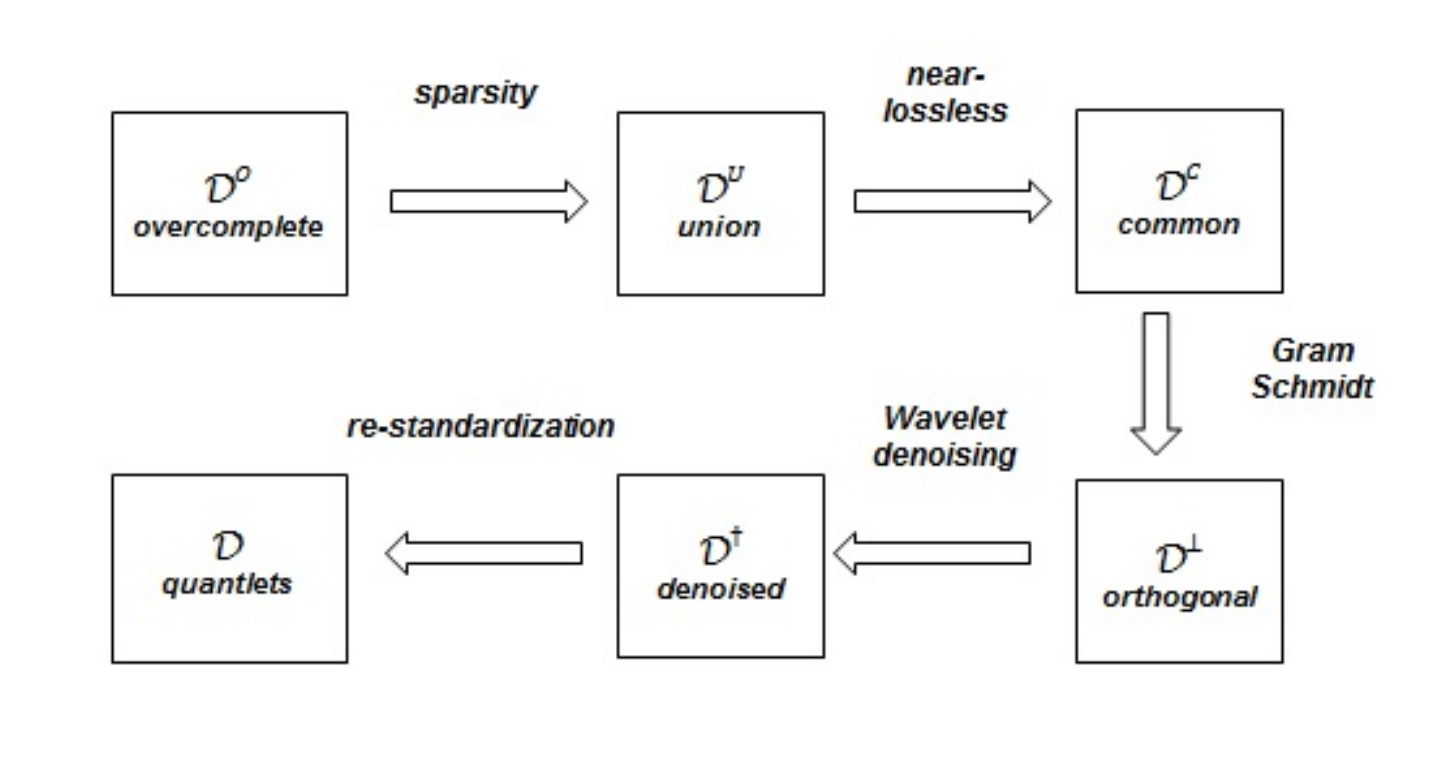}
\vspace{0.5cm} 
\end{figure}

We empirically construct the quantlets for a given data set as a common near-lossless basis that can nearly perfectly represent each subject's empirical quantile function, and then we use these basis functions as building blocks in our quantile functional regression model as described later.  Given a sample of subject-specific empirical quantile functions, we construct a quantlet basis set by the following steps:
\begin{enumerate}
\item Construct an overcomplete dictionary that contains bases spanning the space of Gaussian quantile functions plus a large number of Beta cumulative density functions.  For each subject, use regularization to choose a sparse set among these dictionary elements.  

\item Take the union of all selected dictionary elements across subjects, and find a subset that simultaneously preserves the information in each empirical quantile function to a specified level, as measured by the cross-validated concordance correlation coefficient.

\item Orthogonalize this subset using Gram-Schmidt, apply wavelet denoising to regularize the orthogonal basis functions, and then re-standardize.
\end{enumerate}
We refer to the set of basis functions resulting from this procedure as \textit{quantlets}.   
We describe these steps in detail and then discuss their properties. 
See Figure~\ref{diagram} for an overview of the entire procedure, for which each step is given as follows.

\textbf{Form overcomplete dictionary:}    Suppose that $L^2(\Pi(\mathcal{P}))$ is a Banach space such that $\{ Q: p \in \mathcal{P} \rightarrow \mathbb {R}~~ \text{measurable s.t.}~~ 
\|Q\|_2=\bigr ( \int |Q(p)|^2 d\Pi(p)  \bigl )^{1/2} < \infty \}$, where
$\Pi$ is a uniform density with respect to the Lebesgue measure. We define the first two basis functions to be a constant basis $\xi_1(p)=1$ for $p \in [0,1]$ and standard normal quantile function $\xi_2(p) = \Phi^{-1}(p)$.  These orthonormal bases span the space of all Gaussian quantile functions, with the first coefficient corresponding to the mean and the second coefficient the variance of the distribution.  
We form an overcomplete dictionary that includes these along with a large number of dictionary elements constructed from Beta cumulative density functions (CDF).  The shape of the Beta CDF is able to follow a 
``steep-flat-steep" shape that matches the broad-scale pattern of most empirical quantile functions, so has the potential for efficient representation.

The individual dictionary elements $\xi_k(p)$ are given by
 \begin{equation} \label{p5_phi}
\xi_k(p) =    P_{N^{\perp}} \biggr (    \frac{ F_{\theta_k}(p) -   \mu_{\theta_k} } { \sigma_{\theta_k} }  \biggl ) 
=  P_{N^{\perp}}  \biggr (  \int_{0}^{1} ( I(u \le p)  -\mu_{\theta_k})/\sigma_{\theta_k}dF_{\theta_k}(u)  \biggl ) , 
\end{equation}
where $F_{\theta_k}(p)$ is the CDF of a Beta($\theta_k$) distribution for some positive parameters $\theta_k=\{a_k, b_k\}$, $\mu_{\theta_k}= \int_{0}^1 F_{\theta_k}(u)  du$ and 
$\sigma_{\theta_k}^2= \int_{0}^1 (F_{\theta_k}(u)-\mu_{\theta_k})^2du$ are the centered   
and scaled values of these distributions for standardization, respectively, 
and $P_{N^{\perp}}$
indicates the projection operator onto the orthogonal complement to the Gaussian basis elements $\xi_1(p)$ and $\xi_2(p)$, with $P_{N^{\perp}}\{f(p)\}= f(p) - \xi_1(p) \int_0^1 f(p) \xi_1(p) dp - \xi_2(p) \int_0^1 f(p) \xi_2(p) dp$ .  Put together, the set
$\mathcal{D}^{O}=\{ \xi_1, \xi_2 \} \cup \{  \xi_k:    \theta_k  \in \Theta \}$  comprises an {\it overcomplete dictionary}  family
  on $\Theta \subset \mathbb {R}_{+}^{2}$. 
  In practice, to fix the number of dictionary elements, we choose a grid on the parameter space to obtain $\Theta=\{ \theta_k= (a_k,b_k)  \}_{k=3}^{K^{O}}$ by uniformly sampling on 
 $\Theta \subset (0, J )^{2}$  for some sufficiently large $J$, 
and choosing $K^{O}$ to be a large integer (e.g. we use $K^{O}=12,000$ in this paper).
Details of how to select $\Theta$ can be found in the Supplementary materials.

\textbf{Sparse selection of dictionary elements:} For each $i$, we use regularization via penalized likelihood to obtain a sparse set of dictionary  elements to represent each subject's empirical quantile function.  While other choices of penalty could be used, here we use the Lasso \cite{tibshirani1996regression}, minimizing
 \begin{equation}   \label{p5_lasso}
 \| Q_{i}(p) - \sum_{k \in \mathcal{D}^{O}   } \xi_k(p)Q_{ik}^{O  }\|^2_{2}+
 \lambda_i \sum_{k \in \mathcal{D}^{O}    }\|Q_{ik}^{O}\|_{1}, 
 \end{equation}
for a fixed positive constant $\lambda_i$, where the choice of each $\lambda_i$ 
is determined by cross validation and  $Q_{ik}^{O}$ are basis coefficients for the elements of $\mathcal{D}^{O} $.  
The standardization of the basis functions  ensures they are on a common scale which is important for the regularization method.  
By using the regularization methods, we obtain different sets of  selected {\it dictionary} elements for each 
subject,  denoted by 
$\mathcal{D}_i=\{ \xi_k \in  \mathcal{D}^{O}  :  Q_{ik}^{O} \ne 0 \}$.  Taking the union across subjects, we obtain a unified set of  {\it dictionary} elements denoted by 
 $\mathcal{D}^U= \cup_{i=1}^{n} \mathcal{D}_i$, which we construct to always include the Gaussian basis functions $\xi_1$ and $\xi_2$.

 \textbf{Finding near-lossless common basis:}  
The above sparse selection is done for each subject $i$, however, we would like to use a common 
basis across all subjects to fit the quantile functional regression model.  The unified set of dictionary elements $\mathcal{D}^U$ is likely to be very redundant, with some of the dictionary elements selected for many subjects' empirical quantile functions and many others selected for only a few subjects, and not all necessary.  We would like to  find a common basis set $\mathcal{D}^\mathcal{C}$ that is as sparse as possible while retaining virtually all of the information in the original empirical quantile functions.  
We call such a basis \textit{near-lossless}, which we define more precisely below.

As a measure of losslessness, we use the leave-one-out concordance correlation coefficient (LOOCCC), $\rho_{(i)}$.   This quantifies the ability of a basis set $\mathcal{D}^U_{(i)}$ that has been empirically constructed using all samples except the $i$th one to represent the observed quantile function $Q_i(p)$, with fit measured by the concordance correlation coefficient \cite{lawrence1989concordance}:
 \begin{equation}  \label{p5_rhoi}  
\rho_{(i)}=\frac{ \text{Cov}( Q_i(\cdot), \sum_{k \in   \mathcal{D}^U_{(i)} }  \xi_k(\cdot)  Q_{ik}^{U} ) }
{ \text{Var}( Q_i(\cdot) )  + \text{Var}( \sum_{k \in   \mathcal{D}^U_{(i)} }  \xi_k(\cdot)  Q_{ik}^{U})
+  [ \text{E}(Q_i(\cdot))-  \text{E}(\sum_{k \in   \mathcal{D}^U_{(i)} }  \xi_k(\cdot)  Q_{ik}^{U})]^2    }, 
\end{equation} 
where 
$\text{Cov}$, $\text{Var}$ and $\text{E}$ are taken with respect to ${\Pi}$ and
$Q_{ik}^{U}$ are basis coefficients corresponding to the elements $\xi_k$ contained in the set $\mathcal{D}^U_{(i)}$. 

This measure $\rho_{(i)} \in [0,1]$, with $\rho_{(i)}=1$ indicating the basis set $\mathcal{D}^U_{(i)}$ is sufficiently rich such that there is no loss of information about $Q_i(p)$ in its corresponding projection.  One advantage of this measure over other choices such as mean squared error is that it is scale-free, in the sense that it is invariant to the scale of the quantile functions $Q_i$ and the basis functions $\xi_k$.  Aggregating across subjects, we can compute $\rho^0=\text{min}_i\{\rho_{(i)}\}$ or $\overline{\rho}=\text{mean}_i\{\rho_{(i)}\}$ to summarize the ability of the chosen basis to reconstruct the observed data set in its entirity, with $\overline{\rho}$ the average across all subjects and $\rho^0$ the worst case.  If $\rho^0=1$, we say this basis is \textit{lossless}, and if $\rho^0>1-\epsilon$ for some small $\epsilon$ then we say this basis is \textit{near-lossless}.

\begin{figure}[!htb]
\centering
\caption{Construction of Quantlet Bases.  The concordance correlation for the GBM application: (A) minimum concordance
($\rho^{0}$, red) and average ($\bar\rho$, blue) across samples as function of $K^\mathcal{C}$,
(B) $\rho^{0}$ and $\bar\rho$ for {\it quantlets} basis and principal components, varying with the number of basis coefficients.  
  \label{S5_Figure_qant}}
\includegraphics[height=2.5in,width=5.0in]{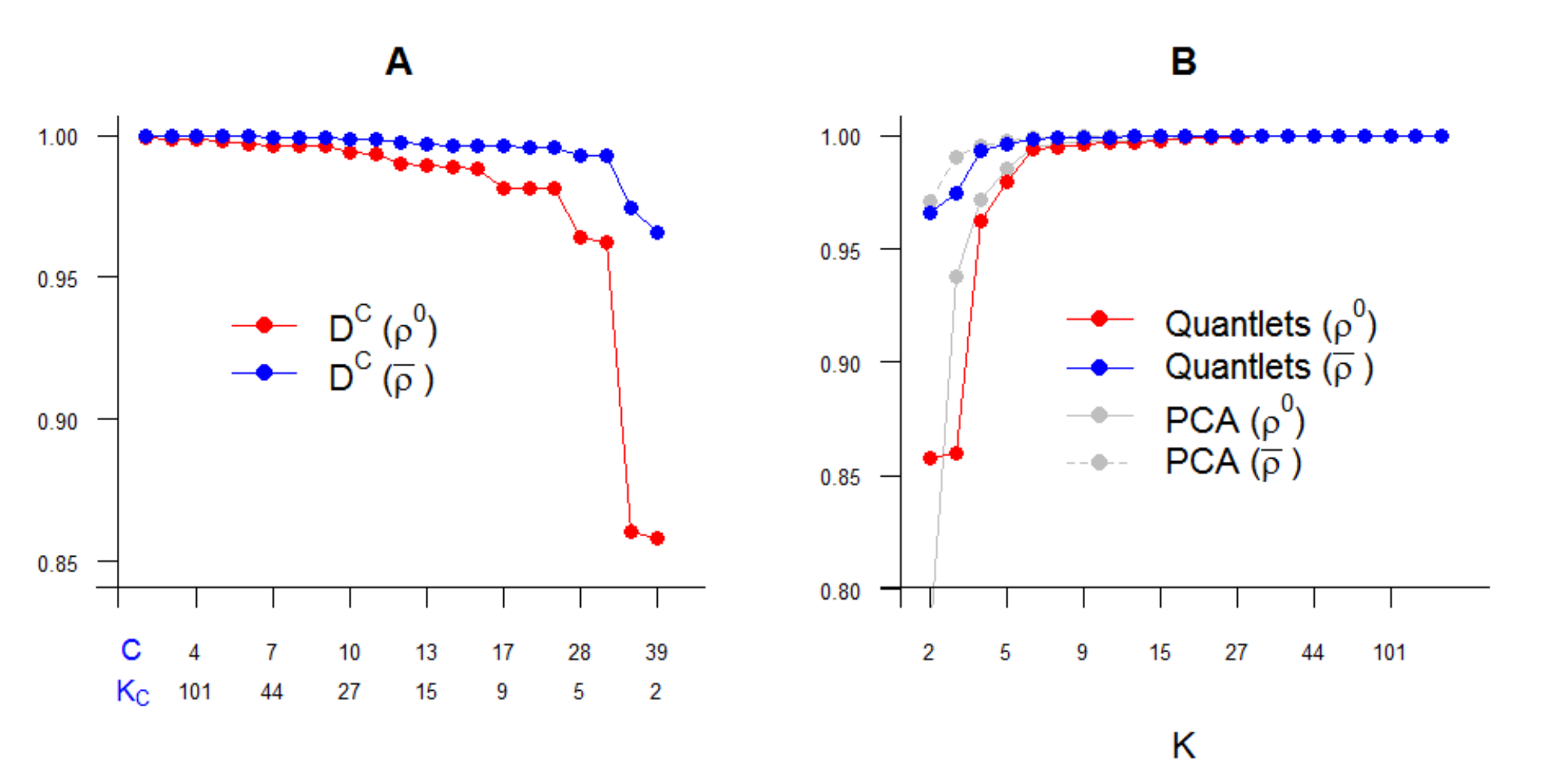}
\end{figure}

To find a sparse yet near-lossless basis set, we define a sequence of reduced basis sets $\{\mathcal{D}^U_{(i)\mathcal{C}}, \mathcal{C}=1, \ldots, n-1\}$ that contain the Gaussian basis functions $\xi_1$ and $\xi_2$ plus all dictionary elements $\xi_k(p)$ that are selected for at least $\mathcal{C}$ of the $n-1$ empirical quantile functions, excluding the $i$th one, i.e. $\mathcal{D}^U_{(i)\mathcal{C}}=\{\xi_k, k:\sum_{i' \ne i = 1}^n I(Q^O_{i'k} \ne 0) \ge \mathcal{C}\}.$ 
We can construct plots of $\rho^0$ or $\overline{\rho}$ vs. $\mathcal{C}$ to choose a value of $\mathcal{C}$ that leads to a sparse basis that can recapitulate the observed data at the desired level of accuracy (as shown below).  Given this choice, we next compute the corresponding reduced basis set using all of the data $\mathcal{D}^\mathcal{C}=\{\xi_k, k:\sum_{i = 1}^n I(Q^O_{ik} \ne 0) \ge \mathcal{C}\}$ containing $K=K_\mathcal{C}$ basis coefficients.  The left panel of Figure \ref{S5_Figure_qant} contains this plot for our GBM data set.  From this, we select $\mathcal{C}=10$ which leads to $K_\mathcal{C}=27$ basis functions, and leads to a near-lossless basis with $\rho^0=0.990$ and $\overline{\rho}=0.998$.

\textbf{Orthogonalization and Denoising:} 
Next, we use Gram-Schmidt to orthogonalize the basis set $\mathcal{D}^\mathcal{C}$ to generate an orthogonalized basis set $\mathcal{D}^{\perp}=\{\psi^\perp_k(p), k=1,\ldots,K\}$, where $\{\psi^\perp_1(\cdot), \psi^\perp_2(\cdot)\}=\{\xi_1(\cdot), \xi_2(\cdot)\}$ comprise the Gaussian basis and $\{\psi^\perp_k(\cdot), k=3,\ldots,K\}$ are orthogonalized basis functions computed from and spanning the same space as the remaining bases in $\mathcal{D}^\mathcal{C}$, indexed in descending order of their total percent variability (total energy) explained for the given data set. 
Specifically, suppose that $Q_{ik}^{\perp}$, $k= 1, \dots, K$ and $i=1, \dots, n$  are the empirical coefficients corresponding to the elements of $\mathcal{D}^\perp$, ordered as in $\mathcal{D}^{\mathcal{C}}$.
We compute the percent total energy for basis $k$ as  $\mathcal{E}_k=\sum_{i=1}^{n}Q_{ik}^{ \perp 2}/\sum_{i=1}^{n}\sum_{k=1}^{K}Q_{ik}^{\perp  2}$, and then relabel $\psi_k, k=3, \ldots, K$ to be in descending order of $\mathcal{E}_k$.

  In practice, we have observed that the first number of orthogonal basis functions are relatively smooth, but the later basis functions can be quite noisy, sometimes with high-frequency oscillations.  As we do not believe these oscillations capture meaningful features of the empirical quantile functions, we regularize the orthogonal basis functions using wavelet denoising to adaptively remove these oscillations.  

 Given a choice of mother wavelet function $\varphi(p)$, wavelets are formulated by the operations of dilation and translation given by  
  $$\varphi_{j,l}(p)=2^{j/2}\varphi(2^j p - l)$$ with integers $j, l$ indicating scale and location, respectively.
We can decompose any arbitrary function $\psi^\perp_k(p) \in L^2(\Pi(\mathcal{P}))$ into the generalized Fourier series as 
 \begin{equation}
 \psi^\perp_k(p) = \sum_{j= -\infty}^{\infty} \sum_{l= -\infty}^{\infty} d_{k,j,l}\varphi_{j,l}(p),
  \end{equation}
where $d_{k,j,l}=\int \psi^\perp_k(p)\varphi_{j,l}(p) dp=\langle \psi^\perp_k, \varphi_{j,l} \rangle$ are the wavelet coefficients corresponding to $\psi^\perp_k$.
Wavelet coefficient $d_{k,j,l}$ describes features of the function $\psi^\perp_k$ at the spatial locations indexed by $l$ and scales indexed by $j$.
 A fast algorithm, the discrete wavelet transform (DWT), can be used to compute these wavelet coefficients in linear time for data sampled on an
 equally spaced grid whose size $L$ is a power of two, yielding a set of $L$ wavelet coefficients, with $L_j$ wavelet coefficients at each of $J$ wavelet scales and $L_0$
 scaling coefficients at the lowest scale. 
 We apply this wavelet transform to the the basis functions $\psi^\perp_k(p)$ sampled on an equally-spaced fine
 grid on $p$, for example using a grid of size $L=2^{10}=1024$ for our GBM data.
 
  Functions can be adaptively denoised by shrinking these wavelet coefficients nonlinearly towards zero \cite{donoho1995wavelet}.  Various shrinkage/thresholding rules can be used to accomplish this, 
such as hard thresholding with a threshold of $\sigma \sqrt{2 \log L}$ introduced by \cite{donoho1995wavelet}, which yields a risk within a log factor of 
the ideal risk.   In that case, the wavelet shrunken and denoised basis function $\psi^{\dagger}_k(p)$ can be constructed as
 \begin{equation}
\psi^{\dagger}_k(p) = \sum_{j= 0}^J \sum_{l= 1}^{L_j} d_{k,j,l}^{\dagger}\varphi_{j,l}(p),
  \end{equation}
such that  $d_{k,j,l}^{\dagger}=d_{k,j,l}$ if $|d_{k,j,l}|>\sigma \sqrt{2 \log L}$ and $d_{k,j,l}^{\dagger}=0$ If $|d_{k,j,l}|\le \sigma \sqrt{2 \log L}$.
When $\sigma$ is unknown, it is often replaced by an empirical estimator that is the median absolute deviation of the wavelet coefficients at the highest frequency level $J$.

  After applying the denoising method to all of the orthogonal basis functions in the set $\mathcal{D}^{\perp}$ to get $\mathcal{D}^\dagger=\{\psi^\dagger_k(p), k=1,\ldots,K\}$, we re-standardize these basis functions  by $\psi_k(p)=(\psi_k^{\dagger}(p) - \mu^\dagger_k)/\sigma^\dagger_k$ for $k=3, \ldots, K$ with $\mu^\dagger_k = \int_0^1 \psi_k^{\dagger}(p) dp$ and $\sigma^\dagger_k=\sqrt{\int_0^1 \{\psi_k^{\dagger}(p)-\mu^\dagger_k\}^2  dp}$ such that $\int_0^1 \psi_k(p) dp=0$ and $\int_0^1 \psi_k(p) \psi_k(p) dp = 1$ for $k=3,\ldots,K$.  

\begin{figure}[!htb]
\centering
\caption{First 16 quantlet basis functions for GBM data set.
  \label{S5_QBE}}
\includegraphics[height=5.3in,width=5.7in]{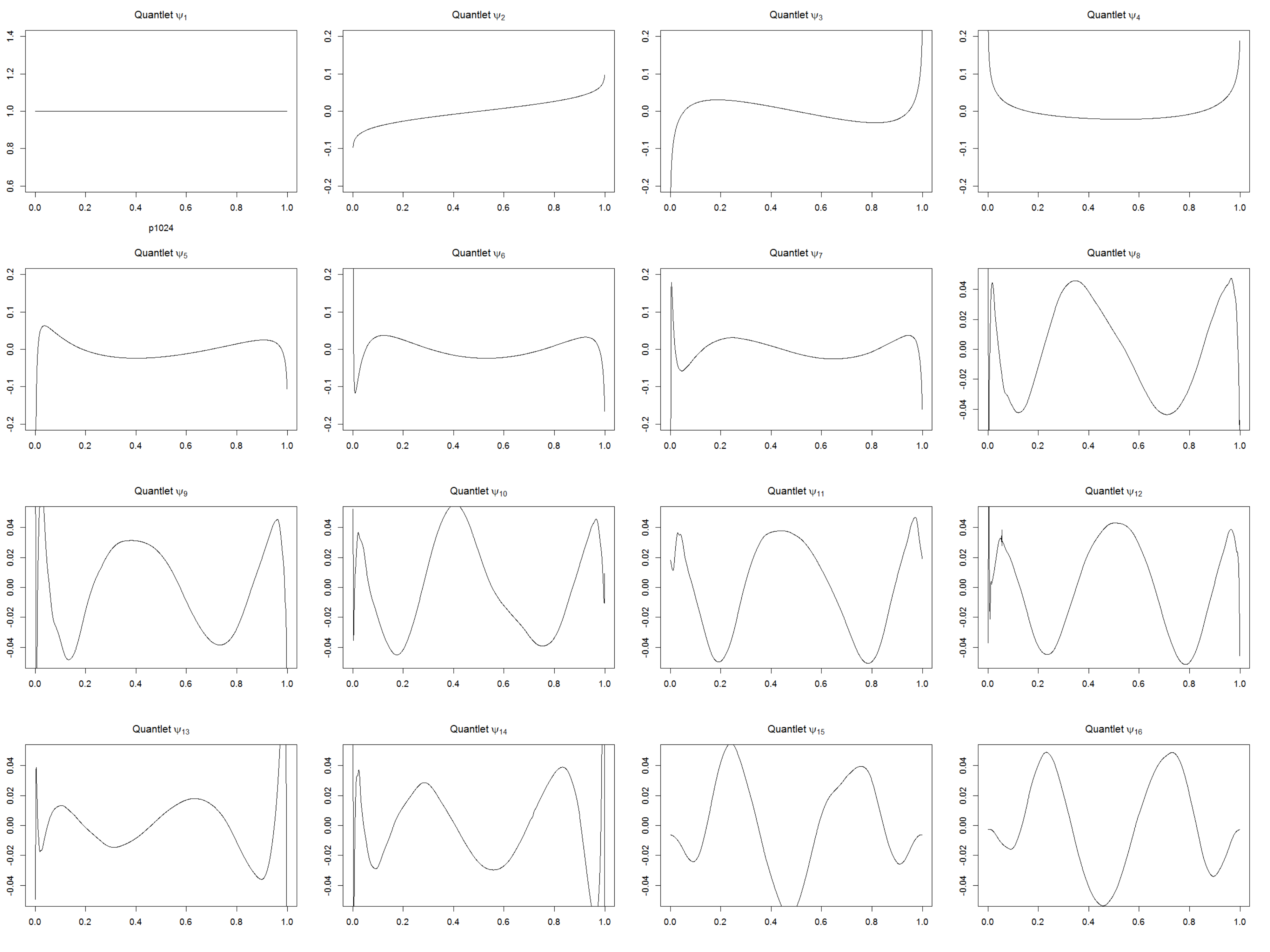}
\vspace{0.5cm} 
\end{figure}

We refer to the resulting basis set $\mathcal{D}=\{\psi_k(p), k=1,\ldots, K\}$ as the \textbf{{\it quantlets}}, which we use as the basis functions in our quantile functional regression modeling.     Figure \ref{S5_QBE} contains the first 16 quantlet basis functions from the GBM data set.

\textbf{Properties of quantlets:}  These quantlets have numerous properties that makes them useful for modeling in our quantile functional regression framework.
\begin{itemize}
\item \textbf{Empirically defined}:  The empirical quantile functions for different applications can have very different features and characteristics.  Given their derivation from the observed data, the quantlets are customized to capture the features underlying the given data set, giving them advantages over pre-specified bases like splines, wavelets, or Fourier series.

\item \textbf{Near-losslessness}: By construction, the set of quantlets are at least \textit{near-lossless} in the sense that the basis is sufficiently rich to almost completely recapitulate the empirical quantile functions $Q_i(p)$.  As a result, we can project the empirical quantiles into the space spanned by the quantlets with negligible error, and thus it is reasonable to consider modeling the quantlet coefficients for the empirical quantile functions as observed data.

\item \textbf{Regularity}: The denoising step tends to remove any wiggles or high frequency noise from the orthogonal basis functions $\psi^\perp_k(p)$, leading to visually pleasing yet adaptive basis functions that are relatively smooth and regular.  We have found these tend to be more regular looking than other empirically determined basis functions like principal components  (compare Figure \ref{S5_QBE} to Supplementary Fig 5).

\item \textbf{Sparsity}: The procedure we have defined to construct the quantlets tends to also produce a basis set that is relatively low dimensional and thus a sparse representation.  We have found these basis functions to have similar sparsity to principal component bases, measured by computing the average LOOCCC $\overline{\rho}$ for quantlets and analogously for principal components (i.e. computing the principal components leaving out the $i$th sample, and then estimating $\rho_{(i)}$ measuring the losslessness of the resulting basis set) -- see Figure \ref{S5_Figure_qant}B and Figure \ref{S2_Figure_1}C.  Using a low dimensional basis enhances the computational speed of our procedure and reduces the uncertainty in the quantile functional regression coefficients $\beta_a(p)$, as can be seen in our sensitivity analyses (Supplementary Table 5).

\item \textbf{Interpretability}:  Unlike principal components, the quantlets have some level of interpretability in that the first two basis functions define the space of all Gaussian quantile functions (see Figure \ref{S5_QBE}).  Also, note that the next two quantlets for the GBM data seem to pick up on fundamental distributional characteristics like the kurtosis and skewness.
For Gaussian data, only the first two basis functions will be needed, while comparing with dimensions $k=3, \ldots, K$ provides a measure of the degree of \textit{non-Gaussianity} in the distribution (see below).
\end{itemize}

\subsection{ \bf{Quantlet-based Modeling in Quantile Functional Regression}}

In this subsection, we describe our quantile functional regression framework based on the  {\it quantlets} basis set.
 We use a \textit{basis transform modeling approach} to fit model (\ref{p5_qfm_i}), which involves
transforming the empirical quantile functional responses to the {\it quantlets} basis space, 
fitting the model in the basis space, and then transforming
the results back to the original function space for interpretation and inference.
 As discussed in the beginning of Section \ref{sec:quantlets}, we use the quantlet basis functions to accommodate correlation across $p$ in the quantile functional regression coefficients $\beta_a(p)$ and residuals $E_i(p)$, which is expected to have advantages over a naive approach that fits independent regression models for a grid of $p$ values.
 
Given the $i$th empirical quantile function $Q_i(p_j)$ evaluated at $p_j=j/(m_i+1), j=1,\ldots, m_i$, constructed from the order statistics $Y_{i(j)}, j=1,\ldots, m_i$, and a quantlet basis set $\mathcal{D}=\{\psi_k(p), k=1,\ldots,K\}$ derived as described in Section \ref{sec:quantlets}, we write a quantlet basis expansion $Q_i(p_j)=\sum_{k=1}^K Q^*_{ik} \psi_k(p_j)$ with $Q^*_{ik}$ being the $k$th empirical quantlet basis function for subject $i$.  For this paper, we will assume that $K<\text{min}_i(m_i)$, with the understanding that $K \ll \text{min}_i(m_i)$ for an extremely large number of applications, including our GBM data.  Extensions of this framework to sparse data settings for which $m_i<K$  for some $i$ are tractable and of interest, but given the length and complexity of this paper and the additional challenges raised by this sparse case, we will leave it to future work.

 With $\bQ_i=[Q_i(p_1), \dots, Q_i(p_{m_i}) ]$ a row vector containing the $i$th empirical quantile function and $\bPsi_i$ a $K \times m_i$ matrix with element $\bPsi_i(k,j)=\psi_k(p_j)$, we can compute the $1 \times K$ vector of empirical quantlet coefficients $\bQ^*_i=[Q^*_{i1}, \ldots, Q^*_{iK}]$ by $\bQ^*_i = \bQ_i \bPsi_i^-$, where $\bPsi_i^-=\bPsi_i^T (\bPsi_i \bPsi_i^T)^{-1}$ is the generalized inverse of $\bPsi_i$.  
 Based on the \textit{near-lossless} property of the quantlets by design, $\bQ^*_i$ contains virtually all of the information in the raw data $\bQ_i$, and thus we model these as our data.

Concatenating $\bQ^*_i$ across the $n$ subjects, we are left with a $n \times K$ matrix $\bQ^*$ that we regress on the covariates $x_{ia}, a=1,\ldots, A$ in the \textit{quantlet space model}
\vspace{-2mm}
\begin{equation}
\bQ^*=\bX \bB^* + \bE^*, \\ \label{eq:quantreg}
\end{equation}
where $\bX$ is an $n \times A$ matrix with $\bX(i,a)=x_{ia}$, $\bB^*$ an $A \times K$ matrix of corresponding quantlet-space regression coefficients, and $\bE^*$ an $n \times K$ matrix of quantlet space residuals.  We can relate this quantlet-space model back to the original quantile functional regression model (\ref{p5_qfm_i}) through the quantlet basis expansions $\beta_a(p)=\sum_{k=1}^K B^*_{ak} \psi_k(p)$ and $E_i(p)=\sum_{k=1}^K E^*_{ik} \psi_k(p)$.

The rows of $\bE^*$ are assumed to be independent and identically distributed mean-zero Gaussians, with $\bE^*_i \sim N(\bzero, \Sigma^*)$.  Here, we assume $\Sigma^*=\text{diag}_k\{\sigma^2_k\}$, which enables us to fit in parallel the models for each column, $\bQ^*_k = \bX \bB^*_k + \bE^*_k, k=1,\ldots,K$, and yet accommodate correlation across $p$ since modeling in the quantlet space induces correlation in the original data space, with the covariance operator for $E_i(p)$ given by $\Sigma(p,p')=\text{cov}\{E_i(p), E_i(p')\} = \bPsi(p) \Sigma^* \bPsi(p')$, where $\bPsi(p)=(\psi_1(p), \ldots, \psi_K(p))^T$.     The empirical nature of the derived quantlets makes this structure well-equipped to capture the key correlations across $p$ in the observed data, as shown for our 
real data set (See Supplementary Figure 9).
If desired, 
one could model $\Sigma^*$ as an unconstrained $K \times K$ matrix, which would provide additional flexibility in the precise form of $\Sigma$ but at a potentially much greater computational cost.

As described below, we fit these models using a Markov chain Monte Carlo-based Bayesian modeling approach, yielding posterior samples for the parameters of the quantlet-space model (\ref{eq:quantreg}).  We project these posterior samples back into the original data space in order to obtain estimates and inference on the quantile functional regression parameters of model (\ref{p5_qfm_i}) on any desired grid of $p$ of size $J$, by $\bB_a=\bB^*_a \bPsi^J$ with $\bPsi^J$ a $K \times J$ matrix with elements $\psi_k(p_j)$.

\subsection{ {\bf Bayesian Modeling Details}}

In order to fit model (\ref{eq:quantreg}) using a Bayesian approach, we need to specify priors on the variance components $\{\sigma^2_k, k=1,\ldots,K\}$ and quantlet-space regression coefficients $\{B^*_{ak}, a=1,\ldots, A, k=1,\ldots,K\}$ comprising the elements of $\bB^*$.

We place a vague proper inverse gamma prior on each diagonal element $\sigma_{k}^2$ given by
    \begin{equation}    \label{p5_prior_G} 
      \sigma_{k}^2 \sim \text{inverse-gamma}( \nu_0/2,  \nu_0/2 ),
  \end{equation}
where $\nu_0$ is some relatively small positive constants. Other relatively vague priors could also be used. 
If one wanted to allow $\Sigma^*$ to be unconstrained, an Inverse Wishart prior could be assumed for the $K \times K$ matrix.

\textbf{Regularization prior for regression coefficients:}  
This model could be fit using vague conjugate priors for the regression coefficients, $\beta^*_{ak} \sim N(0, \tau^2)$ for some extremely large $\tau^2$.  This could be called a \textit{quantlet-no sparse regularization} approach.  It would result in virtually no smoothing of $\beta_a(p)$ relative to the naive (one-$p$-at-a-time) quantile functional regression model, but it would still account for correlation across $p$ in the residual errors, so may have inferential advantages over the naive approach.   We can further improve performance by inducing regularity and smoothness in the quantile functional regression parameters $\beta_a(p)$, which we accomplish through regularization or shrinkage priors, as is customary for Bayesian functional regression models.

Motivated by a belief that the covariate effects should be more regular than the empirical quantile functions themselves, we assume sparsity-inducing priors on the $\beta^*_{ak}$ coefficients.
While many choices could be used, including the Bayesian Lasso
\cite{park2008bayesian}, Horseshoe \cite{carvalho2010horseshoe}, Normal-Gamma \cite{griffin2010inference}, Generalized Double Pareto \cite{armagan2013generalized}, and Dirichlet Laplace \cite{bhattacharya2015dirichlet}, here we use a spike-Gaussian slab \cite{lempers1971posterior,mitchell1988bayesian} distribution.  The spike at 0 induces sparsity while the Gaussian prior applies a roughness penalty.

Motivated by the belief that certain quantlets are \textit{a priori} more likely to be important for representing covariate effects, we partition the set of $K$ quantlet dimensions into $H$ clusters of basis functions, each with their own set of prior hyperparameters.  This allows us, for example, to allow a higher prior probability for certain quantlet dimensions to be important  such as the 
the Gaussian basis levels $\{\psi_1, \psi_2\}$ and the quantlets explaining a high proportion of the relative variability in the empirical quantile functions.  Recalling that quantlets are indexed in descending order of their proportion of relative variability explained, we can group together the Gaussian coefficients as one cluster, and then split the rest sequentially into $H-1$ clusters each containing sets of basis functions whose relative variability explained are of similar order of magnitude  (See Section 2.1 of the supplement for more discussion).

Defining the sets $\mathcal{K}_h, h=1, \ldots, H$ to contain the quantlets $\psi_k$ grouped together within the same cluster, and the index $h_k$ to indicate the cluster of quantlet $k$ with the clustering map, $f(k)=h\equiv h_k$, the prior on $\beta^*_{ak}$ is given by
  \begin{align}    \label{p5_prior_B} 
 & \beta_{ak}^* \equiv\beta_{ah_k,l}^*  \sim  \gamma_{ah_k,l} N(0, \tau^2_{ah_k,l}) + (1-\gamma_{ah_k,l})I_{0}       \\
 &\gamma_{ah_k,l}  \sim \text{Bernoulli}(\pi_{ah}) ,      \notag 
  \end{align}
  where $I_{0}$ is a point mass distribution at zero, and $\gamma_{ah_k,l} \equiv \gamma_{ak}$ is an indicator of whether the $k$th quantlet basis coefficient is important for representing the effect for the $a$th covariate within the $h$ cluster,   when $h_k$ is the $l^{\text{th}}$ component in the $\mathcal K_{h}$ cluster, 
  $l \in \mathcal{K}_h = \{1,\dots,|\mathcal{K}_h| \}$.    
 The hyperparameter $\pi_{ah}$ indicates the prior probability that a quantlet coefficient in set 
  $\mathcal{K}_h$ is important, and $\tau^2_{ah_k,l}$ the prior variance, and regularization factor, for coefficient $B^*_{ak}$ conditional on it being chosen as important.  
  
The parameters $\pi_{ah}$ and $\tau^2_{ah_k,l}$ can be estimated using an empirical Bayes method following a similar procedure as in  \cite{morris2006wavelet}, as detailed in the Section 2.2 of the supplementary materials, or alternatively could be given hyperpriors.

\subsection{ {\bf MCMC Sampling}}
We fit the quantlet space model (\ref{eq:quantreg}) using Markov chain Monte Carlo (MCMC).
  
Let ${\bf Q}_k^*$ and ${\bB}_{k}^{\ast}$  
  be the $k$th column vector of ${\bQ}^{\ast}$ and ${\bB}^{\ast}$, respectively.
  For each quantlet basis $k=1,\ldots,K$, we sample the $a$th covariate effect from $f(\beta_{ak}^{\ast} | {\bf Q}^{\ast}, {\bB}_{(-a)k}^{\ast},\sigma^2_k)$, where ${\bB}_{(-a)k}^{\ast}$ is a vector of length $A-1$ containing all covariate effects except the $a$th of $\bB^*$  in  model (\ref{eq:quantreg}) for the $k$th quantlet coefficient.
 We repeat this procedure for all covariates,  $a=1, \dots, A$ and quantlet basis function $k=1,\ldots,K$.
This distribution is a mixture of  a point mass at zero and a normal distribution, with normal mixture proportion $\alpha_{ak}$ and the mean and variances of the normal distribution $\mu_{ak}$ and $v_{ak}$ given by 
   \begin{align*}
   \beta_{ak}^{\ast} \equiv\beta_{ah_k,l}^*  \sim 
   \alpha_{ah_k,l}
    N( \mu_{ah_k,l}, v_{ah_k,l}) +
  (1- \alpha_{ah_k,l})I_{0}
\end{align*} 
where  $\alpha_{ah_k,l}$, $\mu_{ah_k,l}$ and  $v_{ah_k,l}$ are given by 
   \begin{align*}
&\alpha_{ah_k,l}
=     \text{P}( \gamma_{ah_k,l} =1 |  {\bf Q}^{\ast}_k, {\bB}_{(-a)k}^{\ast},\sigma^2_k)   
=  \widehat{O}_{ah_k,l}/( \widehat{O}_{ah_k,l}+1)\\
&\mu_{ah_k,l}= \widehat{\beta}_{ah_k,l}^{\ast}(1+ V_{ah_k,l}/\tau_{ah_k,l})^{-1} \\
&v_{ah_k,l}= V_{ah_k,l}(1+ V_{ah_k,l}/\tau_{ah_k,l} )^{-1} .  \notag \\
& \widehat{O}_{ah_k,l}=  \frac{\hat \pi_{ah} }{1 - \hat \pi_{ah} }(1 + V_{ah_k,l}/\tau_{ah_k,l} )^{-1/2}
  \exp{\biggl \{ 
  \frac{1}{2}\xi_{ah_k,j}^2 \frac{ V_{ah_k,l}/\tau_{ah_k,l} }{1+V_{ah_k,l}/\tau_{ah_k,l} } 
  \biggr \}} \\
 & \xi_{ah_k,l}= \widehat{\beta}_{ah_k,l}^* /V_{ah_k,l}^{1/2} \\
  & V_{ah_k,l}=\left(\sum_{i=1}^{n} x_{ia}^2/\sigma_{k}^2\right)^{-1} 
\end{align*} 
For each quantlet basis $k=1,\ldots,K$, we sample $\sigma_k^2$ from 
its complete conditional
 \begin{align*}
\text{P}( \sigma_{k}^{2}|{\bB}^{\ast}_k, {\bf Q}^{\ast}_k,  {\bX})
\sim  \text{Inverse Gamma}\{  (\nu_0+n)/2,
(\nu_0+ \text{SSE}({\bB}^{\ast}_k) )/2\},
\end{align*}
where $\text{SSE}({\bB}^{\ast}_k)= {\bf Q}_k^{\ast  T}({\bf I} -{\bX}
( {\bX}^{T}{\bX}   )^{-1}{\bX}^{T} ){\bf Q}_k^{\ast}$.

  \subsection{ \bf{Posterior Inference}} \label{sec:BayesInference}

The aforementioned MCMC algorithm produces posterior samples for all quantities in  
the {\it quantlet space}. 
These posterior samples are transformed back to the {\it data space} using 
${\beta}_a^{(m)}(p) = \sum_{k=1}^{K} B_{ak}^{ \ast (m)}\psi_k(p), m=1,\ldots,M$ where $M$ is the number of MCMC samples after burn in and thinning.
From these posterior samples, various Bayesian inferential quantities can be computed, including point wise and joint credible bands, global Bayesian p-values, and multiplicity-adjusted probability scores, as detailed below.  These can be computed for $\beta_a(p)$ itself or any transformation, functional, or contrast involving these parameters.

\textbf{Point and joint credible bands:}  
Pointwise credible intervals for $\beta_a(p)$ can be constructed for each $p$ by simply taking the $\alpha/2$ and $1-\alpha/2$ quantiles of the posterior samples.  Use of these local bands for inference does not control for multiple testing, however.  Joint credible bands have global properties, with  the $100(1-\alpha)\%$ joint credible bands for $\beta_a(p)$ satisfying $  \text{P}(L(p) \le   \beta_a(p) \le U(p) ~~ \forall p \in \mathcal{P} ) \ge 1-\alpha$.  Using a strategy as described in  \cite{ruppert2003semiparametric},  we can construct joint bands by
   \begin{align} \label{p5_jointci}
J_{a,\alpha}(p)= \hat{\beta}_a(p) \pm q_{(1-\alpha)}  \bigl [ \widehat{\text{St.Dev}}\{\hat \beta_a(p)\} 
\bigr ],
\end{align} 
where $\hat \beta_a(p)$ and $\widehat{\text{St.Dev}}\{\hat \beta_a(p)\}$ are the mean and standard deviation
for each fixed $p$ taken over all MCMC samples. 
Here the variable $q_{(1-\alpha)}$ is the $(1-\alpha)$
quantile taken over all MCMC samples of the quantity
   \begin{align*}
Z_a^{(m)}= \max_{p \in \mathcal{P}}\left| \frac{\beta_a^{(m)}(p)-\hat \beta_a(p)}{\widehat{\text{St.Dev}}\{\hat  \beta_a(p)\} }  \right|.
\end{align*}

\textbf{SimBaS and GBPV:}  
Following \cite{meyer2015bayesian}
we can construct  $J_{a,\alpha}(p)$ for multiple levels of $\alpha$ and
determine for each $p$ the minimum $\alpha$ such that $0$ is excluded 
from the joint credible band, which we call {\it Simultaneous Band Scores (SimBaS)}, $\text{P}_{a,SimBaS}(p)=\min{\{\alpha: 0\not\in  J_{\alpha}(p) \}}$, which can be directly estimated by
   \begin{align*}
\text{P}_{a,SimBaS}(p)=  M^{-1}\sum_{m=1}^{M}I\biggl \{ 
\biggl |
\frac{ \hat\beta_a(p) }{\widehat{\text{St.Dev}}\{\hat \beta_a(p)\}}
\biggr |
\le Z_a^{(m)}
\biggr \}.  
\end{align*}  
These can be used as
local probability scores that have global properties, effectively adjusting for multiple testing.  For example, we can flag all $\{p:\text{P}_{a,SimBaS}(p)<\alpha\}$ as significant.
From these we can compute $P_{a,Bayes}=$min$_p\{P_{a,SimBaS}(p)\}$, which we call {\it global
Bayesian p-values} (GBPV) such that we  reject the global hypothesis that $\beta_a(p) \equiv 0$ whenever 
$P_{a,Bayes}<\alpha$.

\textbf{Probability score and moments:}  
As mentioned in Section \ref{sec:QF}, distributional moments can be constructed as straightforward
functions of the quantile function, and thus from posterior samples of quantile functional regression
parameters one can construct posterior samples of these moments for various levels of covariates
$\bX$.
Denoting ${\beta}^{(m)}(p)=({\beta}_1^{(m)}(p), \dots, {\beta}_A^{(m)}(p))^{T}$ 
for each MCMC sample $m=1,\ldots,M$, posterior samples of distributional moments conditional
on $\bX$ are given by
 \begin{align}  \label{conmoments}
&{ \mu}^{(m)}_{\bX}  =  \int_{0}^{1} \bX^{T}\beta^{(m)}(p)  dp   \notag \\ 
&{\sigma}^{2(m)}_{\bX} =  \int_{0}^{1}(\bX^{T}\beta^{(m)}(p) - { \mu}^{(m)}_{\bX}  )^2 dp,   \notag \\ 
&{\xi}^{(m)}_{\bX}=  \int_{0}^{1}(\bX^{T}\beta^{(m)}(p) -{ \mu}^{(m)}_{\bX}  )^3/{\sigma}^{3(m)}_{\bX} dp, ~~\text{and}\notag \\  
&{\varphi}^{(m)}_{\bX}=   \int_{0}^{1}(\bX^{T}\beta^{(m)}(p) - { \mu}^{(m)}_{\bX} )^4/{\sigma}^{4(m)}_{\bX} dp.  
\end{align}
The conditional expectations of other basic statistics are similarly derived.   
We can construct posterior probability scores to assess differences of moments between groups or specific levels of continuous covariates as follows.  For each posterior sample, we compute the appropriate moment from the formulas in (\ref{conmoments}) for two covariate levels, $\bX_1$ and $\bX_2$, and compute the difference, e.g. for the mean $\Delta_m = \mu_{1m} - \mu_{2m}$.  Then, we 
define the posterior probability score for the comparison as:
$$P_{\mu_1-\mu_2}=2\min{\{  M^{-1} \sum_{m=1}^{M}I(\Delta_m>0) \},  M^{-1} \sum_{m=1}^{M}I(\Delta_m<0) \}}$$
In assessing a dichotomous covariate $x_{a}$, we compare $x_a=0$ and $x_a=1$ while holding all other covariates at the mean, while when assessing a continuous covariate we compute differences for two extreme values of $x_a$, with the corresponding probability scores for the respective moments denoted $P_{a,\mu}$, $P_{a,\sigma}$, $P_{a,\xi}$, or $P_{a,\varphi}$.

\textbf{Summarizing Gaussianity:}  
As mentioned above, the first two quantlets form a complete basis for the space of Gaussian quantile functions, so by comparing the first two coefficients to the remainder one can obtain a rough measure of ``Gaussianity'' of the predicted
distribution for a given set of covariates $\mathbf{X}$.  One measure that can be computed is $\sum_{k=1}^2 (\mathbf{X} \hat{\beta}_{ak})^2/\sum_{k=1}^K (\mathbf{X}\hat{\beta}_{ak})^2$, which will be on $[0,1]$, with a value of 1 precisely when the predicted quantile function is completely determined by the first two (Gaussian) bases.

\textbf{Predicted PDF and CDF:}  
To some researchers, distribution functions or probability density functions are more intuitive than quantile functions, and given their one-to-one relationship, it is possible to construct CDF or PDFs from the posterior samples as follows.  CDFs can be constructed by simply plotting $p$ vs. $\text{E}\{\hat{Q}(p)|\mathbf{X},\mathbf{Y}\}$, and given posterior samples of the predicted quantile functions on an equally spaced grid $0<p_1, \ldots, p_J<1$, one can estimate predicted pdf for a set of covariates as described in Section 2 of the supplement. 

 \noindent Following is our recommended sequence of Bayesian inferential procedures.
 \begin{enumerate}[1.]
 \item Compute the global Bayesian p-value $P_{a,Bayes}$ for each predictor or contrast. 
\item For any covariates for which $P_{a,Bayes}<\alpha$, characterize the differences:
\vspace{-0.4cm}
 \begin{enumerate}[2a.]
\item Flag which probability grid points $p$ are different using $P_{SimBas}(p)<\alpha$.
\item  Compute moments; assess which moments differ according to the covariates.
\item  Assess whether the degree of Gaussianity appears to differ across covariates.
 \end{enumerate}
 \vspace{-0.4cm}
\item If desired, compute the predicted densities or CDFs for any set of covariates.
 \end{enumerate}

\section{ {\bf Simulation Study}}

We conducted a simulation study to evaluate the performance of the quantile functional modeling framework and the use of quantlet basis functions.

We generated random samples for four groups of subjects whose mean quantile function was assumed to be from a skew normal distribution
 \begin{equation}
f(x)=\frac{2}{\omega} \phi \biggl ( \frac{x-\eta}{\omega} \biggr )  
\Phi \biggl ( \alpha   \biggl ( \frac{x-\eta}{\omega} \biggr )  \biggr )  
  \label{skewnormal} 
 \end{equation}
with the respective values of $(\eta, \omega, \alpha)$  being
$(1, 5, 0)$, $(3, 5, 0)$, $(1, 6.5, 0)$, and $(9.11, 7.89, -4)$, which correspond to a $N(1,5)$, $ N(3,5), N(1,6.5)$, and a skewed normal with mean $1$, variance $5$, and skewness $-0.78$.    
Panels A and E of Figure \ref{S2_Figure_1} below show the densities
and quantile functions, respectively, corresponding to these distributions.
 
For each group $j=1,\ldots, 4$, we generated the random process $Q_{ij}(p)$ for $i=1,\ldots n$ subjects, taking 1024 samples from the corresponding skewed normal distribution, with $p\in \mathcal{P} = [1/1025, \ldots, 1024/1025]$, and some correlated
noise $\epsilon_{ij}(p)$ added to allow some random biological variability in the individual subjects' distributions.  That is, $Y_{ij}(p) = \beta_j(p)+\epsilon_{ij}(p)$,  where $\epsilon_{ij}(p)$ follows an Ornstein-Uhlenbeck process such that  $\text{Cov}(\epsilon_{ij}(p), \epsilon_{ij}(p') )=0.9^{|p-p'|}$.

After constructing the empirical quantile function $Q_{ij}(p)$ by reordering $Y_{ij}(p)$ in $p$, the quantile functional regression model we fit to these data was
 \begin{equation}
 Q_{ij}(p)= \sum_{a=1}^4 X_{ija} \beta_a(p) + \epsilon_{ij}(p),
  \label{mgroups} 
 \end{equation}
with covariates defined such that $X_{ij1}=1$ is for the intercept and $X_{ija}=\delta_{j=a}$ for $a=2, 3, 4$ group indicators for groups 2-4.  Note that with this parameterization, the means of the four groups are, respectively,
$\beta_1(p)$, $\beta_1(p)+\beta_2(p)$, $\beta_1(p)+\beta_3(p)$, and $\beta_1(p)+\beta_4(p)$, and by construction
$\beta_2(p)$ represents a location offset, $\beta_3(p)$ a scale offset, and $\beta_4(p)$ a skewness offset.  Panel E of Figure \ref{S2_Figure_1} displays the true mean quantiles for each group and panel F the true values for these quantile functional regression coefficients.

 \begin{figure}[!htb]
 \caption{ Simulated data of four groups and their {\it quantlet} representations:  (A) density functions of the population, (B) the near-lossless criterion varying with the different number of basis functions, (C) the concordance correlation varying with the cumulative number of the {\it quantlets}, and compared with principal components  (D) the relation between empirical quantile functions and {\it quantlet} fits, (E) mean quantile functions by group and (F) quantile functional regression coefficient estimates.
  \label{S2_Figure_1}}
\centering
\includegraphics[height=3.5in,width=5.5in]{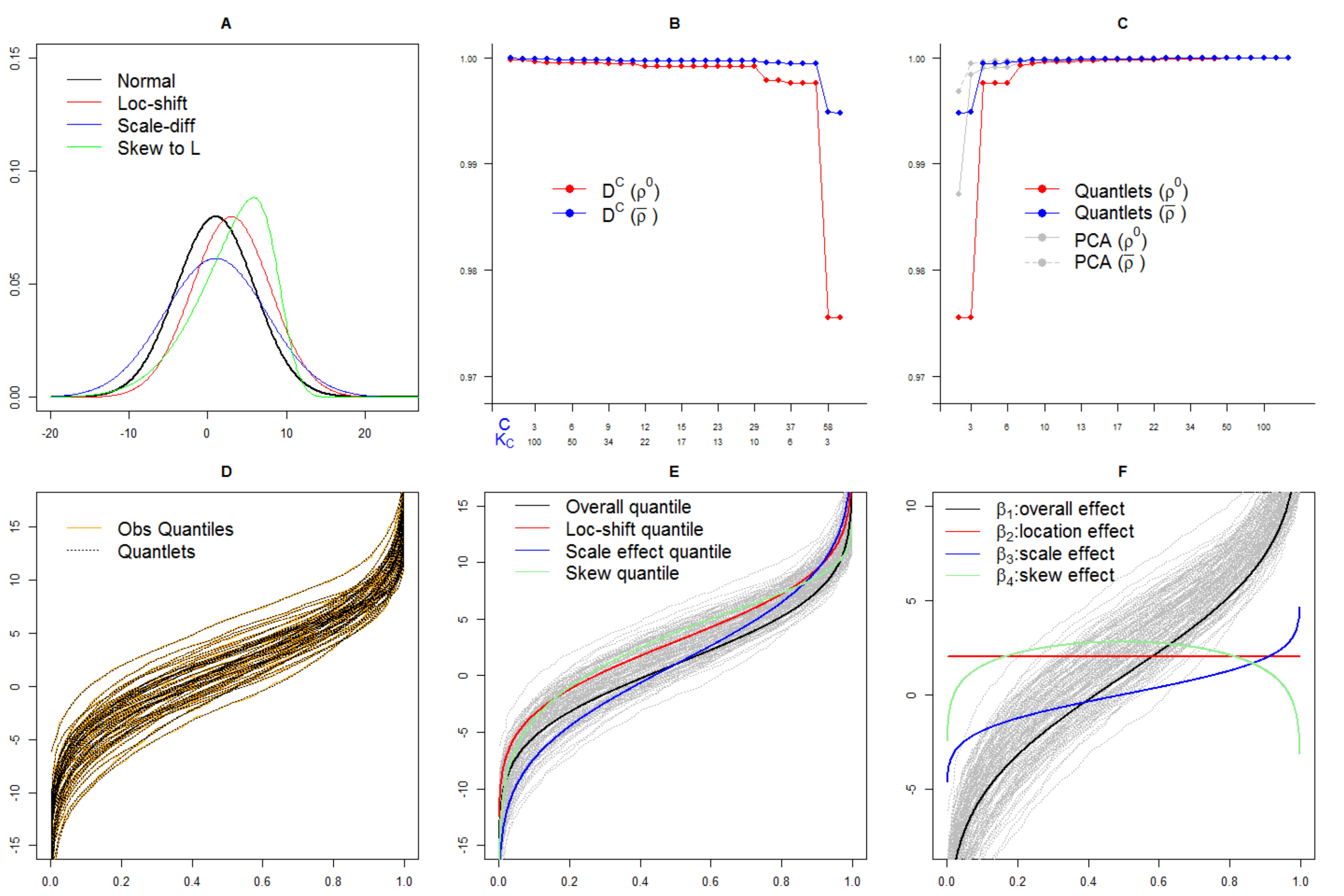}
\end{figure}

We constructed a quantlet basis set for this data set as described above, with some results summarized in panels B, C, and D of Figure \ref{S2_Figure_1}.  
The union set $\mathcal{D}^{U}= \cup_{i=1}^{n} \mathcal{D}_i$ included $2,868$ basis functions, and 
we chose a {\it common} set, $\mathcal{D}^{\mathcal{C}}$, that retained $10$ basis functions, which
resulted in a near-lossless basis set with $\rho^0=0.997$ (see $K_\mathcal{C}=10$ in  panel B).  After orthogonalization, denoising, and re-standardization, the
set of quantlets had sparsity properties similar to principal components (see panel C),
 and the fitted {\it quantlet} projection almost perfectly coincided with the observed data for all of the empirical quantile functions (panel D).  Supplemental Figure 4 contains a plot of these $10$ quantlet basis functions.

We applied several different approaches to these data: 
(A) naive quantile regression method (separate classical quantile regressions for each $p$ 
 by using  \textit{rq} function in {\it quantreg} R package \cite{koenker2005quantile}
),
 (B) naive quantile functional regression approach (separate functional regressions for each subject-specific quantile $p$),
    (C) principal components method (quantile functional regression using PCs as basis functions),
(D) {\it quantlet} without sparse regularization, (E) {\it quantlet} with sparse regularization, and 
(F) Gaussian model (quantlet approach but keeping only the first two coefficients).
The naive quantile regression method (A) ignores all intrasubject correlation in the data and estimates the \textit{population} quantile conditional on covariates, not the \textit{subject-specific} quantile conditional on covariates desired in this quantile functional regression setting, but it is included here since some researchers may choose this approach for these data and we wanted to demonstrate that it is not a good idea..  
In each case, the MCMC was run for $2,000$ iterations, keeping every one after a burn-in of $200$.
The results are shown in Supplementary Figure 8. 
 We compared the methods in terms of the area within the joint credible region and the 
corresponding integrated coverage rate, defined respectively as 
$\mathcal{A}(a)= \int_{0}^{1} | J_{a,\alpha}^{upper}(p) -  J_{a,\alpha}^{lower}(p) |^2 dp$ and
$ \mathcal{C}(a)= \int_{0}^{1} I( J_{a,\alpha}^{lower}(p) \le {\bbeta}_a(p)
  \le J_{a,\alpha}^{upper}(p) ) dp$,
where $J_{a,\alpha}^{upper}(p)$ and $J_{a,\alpha}^{lower}(p)$ are
the upper and lower joint credible bands, respectively.

To investigate the degree of monotonicity afforded by the model, we constructed predicted quantile functions for a broad range of covariate values, and computed the degree of $\epsilon$-monotonicity, defined to be $P^\mathcal{M}_\epsilon(X)=\int_0^1 I[\widehat{Q}(p|X)-\text{max}_{p'<p}\{\widehat{Q}(p|X)\}>\epsilon] dp$ for some $\epsilon$ considered negligibly small in the context of the scale of $Y$ in the current data set.  
 We report the empirical rates of the $\epsilon$-monotonicity as $1-n^{-1}\sum_{i=1}^{n}P^\mathcal{M}_\epsilon(X_i)$.  This empirical summary measure can be used to assess if a given model produces predictors with significant non-monotonicities across $p$ or not.

Table \ref{S2_CI} reports $\mathcal{A}(a)$ and  $\mathcal{C}(a)$ for all quantile functional coefficients.  Methods A-E all had good coverage properties, but use of the basis
functions in modeling (C, D, E) clearly led to tighter joint credible bands than the naive quantile regression and naive quantile function regression methods that did not borrow strength across $p$, as expected, and the use of sparse regularization (E) led to tighter bands than the quantlet method with no shrinkage (D).  Supplementary Figure 8 demonstrates the wiggliness and extremely wide joint credible bands of the naive methods.
Note also that for the coefficient with significant skewness $\beta_4(p)$, the Gaussian model (F) had extremely poor coverage, while for the coefficients corresponding to the Gaussian groups, the quantlet model (E) had performance no worse than the Gaussian method.   This is encouraging, suggesting that when the quantile functions are Gaussian there is not much loss of efficiency from using a richer quantlet basis set.  

Supplementary Figure~10 
depicts the simultaneous band scores $P_{SimBaS}(p)$ for the two contrast functions associated with the scale effect $\beta_3(p)$ and skewness effect
$\beta_4(p)$, with regions of $p$ for which $P_{SimBaS}(p)<0.05$ are flagged as significantly different.  As seen in  Supplementary Figure 10, 
we expect to flag the tails in the scale effect and a broad region in the middle and in the extreme tails for the skewness effect.  Note how the quantlet method with sparse regularization (E) flagged a larger set of regions than the other approaches, especially (B).  In all cases, the global adjusted Bayesian p-values $\text{P}_{Bayes}=\min{\{\text{P}_{map}(p)\}}$ were less than $0.0005$; hence, 
the null hypothesis  $\beta_a(p)\equiv 0$ was rejected in all models.

\begin{table}
\caption{Simulation Results: Area and coverage for the joint 95\% confidence intervals:
 (A) naive quantile regression approach,
(B) naive quantile functional regression approach,
(C) principal component method, (D) {\it quantlet} space without sparse regularization, 
(E) {\it quantlet} space with sparse regularization, and (F) Gaussian {\it quantlet} space approach.
      \label{S2_CI}}
\begin{center}
\resizebox{0.98\columnwidth}{!}{  
\begin{tabular}{ c | c c c c  c c } \hline
 $\text{ {\bf Type} }$     
  & $\text{ {\bf A} }$ &  $\text{{\bf B}}$ & $ \text{{\bf C}}$  & $ \text{{\bf D}}$ &$ \text{{\bf E}}$  &$ \text{{\bf F}}$   \\ \hline
${\beta_1(p)}$ 	
&2.448 (1.000) &	1.603 (1.000)	&	1.092 (0.999)	&	1.186  (1.000)  & 1.069 (1.000)  & 1.071 (1.000)  \\
${\beta_2(p)}$ 	
& 3.487 (1.000) &	2.246 (1.000)	&	1.551 (1.000)	&	1.706 (1.000)  & 1.465 (1.000) & 1.551 (1.000)   \\
${\beta_3(p)}$ 
& 3.581 (1.000)
&	2.242 (1.000)	&	1.599 (1.000)	&	1.717  (1.000)   & 1.457 (1.000) & 1.599 (1.000)  \\ 	
${\beta_4(p)}$ 	& 3.658 (1.000)
&	2.281 (1.000)	&	1.583 (1.000)	&	1.651  (1.000)  & 1.499 (1.000)  & 1.520 (0.421)  \\
	\hline	
\end{tabular}
}
\end{center}
\end{table}      
\begin{table}

\caption{Simulation: Testing for conditional moment statistics in simulation:
 (A) naive quantile regression approach,
(B) naive quantile functional regression approach,
(C) principal component method, (D) {\it quantlet} space without sparse regularization, 
(E) {\it quantlet} space with sparse regularization, (F) Gaussian {\it quantlet} space approach,
   and  (G) feature extraction approach,
  where the values in this table are the posterior probability scores derived by its corresponding method for each test (the first column).
      \label{S2_Test}   
    }
\begin{center}
\resizebox{0.72\columnwidth}{!}{   
\begin{tabular}{c   c| c c c  c c c c} \hline
  $\text{{\bf H}}_0$    &$\text{ {\bf True} }$   
 & $\text{ {\bf A} }$ &  $\text{{\bf B}}$ & $ \text{{\bf C}}$    & $ \text{{\bf D}}$   
  & $ \text{{\bf E}}$        & $ \text{{\bf F}}$             & $ \text{{\bf G}}$       
    \\ \hline
$\mu_{1}=\mu_{3}$   &	$\mu_{1}=\mu_{3}$	 &0.205
&	0.001	&	0.193	&	0.211	&	0.217	&	0.212	&	0.205 \\
$\mu_{2}=\mu_{4}$   &	$\mu_{2}=\mu_{4}$	 &0.438
&	0.001	&	0.447	&	0.465	&	0.445	&	0.462	&	0.438 \\
$\sigma_{1}=\sigma_{3}$   &	$\sigma_{1}\neq\sigma_{3}$	 	 &0.001
&	0.001	&	0.001	&	0.001	&	0.001	&	0.001	&	0.001  	\\
$\sigma_{2}=\sigma_{4}$   &	$\sigma_{2}=\sigma_{4}$		&0.187
&	0.002	&	0.420	&	0.334	&	0.331	&	0.016	&	0.187  	\\
$\xi_{1}=\xi_{3}$   &	$\xi_{1}=\xi_{3}$	 &0.389  
	&	0.374	&	0.498	&	0.488	&	0.479	&	0.493	&	0.389  \\
$\xi_{2}=\xi_{4}$   &	$\xi_{2}\neq\xi_{4}$	 &0.001	
&	0.001	&	0.001	&	0.001	&	0.001	&	0.505	&	0.001 \\
\hline
\end{tabular}
}
\end{center}
\end{table}      
We computed posterior probabillity scores to compare the mean, standard deviation, and skewness for each pair of distributions (Table \ref{S2_Test}), and
Supplemental Table 2 
contains the posterior means and credible intervals for each summary.  We see that the basis function methods (C-E) all flagged the
correct differences, while the naive quantile functional regression approach (B) had major type I error problems in the moment tests and the 
Gaussian method (F) unsurprisingly was unable
to detect differences in skewness.  As an additional comparison, we also applied the so-called {\it feature extraction} approach (G), which involved first
computing the moments from the set of values for each subject and then performing statistical test comparing these across the groups.  Encouragingly, we found
these results were near identical to those found using our quantile functional regression with quantlets (E), suggesting that our unified functional
modeling approach does not lose power relative to feature extraction approaches when the distributional differences are indeed contained in the
moments. 

Constructing predicted quantile functions for a wide range of predictors and assessing $\epsilon$-monotonicity, we found that the all predicted quantile functions from the quantlet-based methods were monotone, while the naive quantile functional regression method had $\epsilon$-monotonicity of 25.8\% and 96.8\% for $\epsilon=0.001$ and $0.01$, respectively, demonstrating that quantlet basis functions encouraged the predicted quantile functions to be monotone in $p$.

\section{{\bf Quantile Functional Regression Analysis of GBM Data}}

Glioblastoma multiforme (GBM) is the most common and most aggressive form of primary brain cancer.  Studying GBM is difficult in that  the cause of most cases is unclear, there is no known way to prevent the disease, and most people diagnosed with GBM survive only  12 to 15 months, with less than $3\%$ to $5\%$ surviving longer than five years \cite{tutt2011glioblastoma}. Most GBM diagnoses are made by medical imaging such as  computed tomography, magnetic resonance imaging (MRI), and positron emission tomography. MRI is frequently chosen because it offers a wide range of high-resolution image contrast that can   serve as indicators for clinical decision making or for tumor progression in GBM studies. 
A GBM tumor, which usually originates from a single cell,  demonstrates heterogeneous physiological and morphological features  as it proliferates \cite{marusyk2012intra}.  Those heterogeneous features make it difficult to predict treatment impacts and outcomes for patients with GBM.  As pointed out by \cite{felipe2013cancer}, investigating tumor heterogeneity is critical in cancer research since inter/intra-tumor differences have stymied the systematic development of targeted therapies for cancer patients. It is of scientific interest to identify the association between characteristics originating from tumor heterogeneity and  clinical measurements within an integrated model.  Thus, our primary goal here is to assess how variability in image intensities in the tumor relate with various clinical, demographic, and genetic factors.   

In our GBM case study, radiologic images consisting of pre-surgical T1-weighted post contrast MRI sequences from 64 patients were obtained from from the Cancer Imaging Archive (cancerimagingarchive.net), along with measurements of certain covariates, including sex (21 females, 43 males), age (mean 56.5 years), DDIT3 gene mutation (6 yes, 58 no), EGFR gene mutation (24 yes, 40 no), GBM subtype (30 mesenchymal, 34 other), and survival status (25 less than 12 months, 39 greater than or equal to 12 months).  The images were preprocessed according to \cite{saha2016demarcate}, from which we extracted the set of $m_i$ pixel intensities within the demarcated tumor region for each patient $i=1,\ldots,n=64$.  The number of pixels within the tumor ranged from 371 to 3421.

We sorted the pixel intensities for each patient, yielding an empirical quantile function $Q_i(p_{ij})$ on a grid of observational points $p_{ij}=j/(m_i+1), j=1, \ldots, m_i$.  We related these to the clinical, demographic, and genetic covariates using the following quantile functional regression model:
 \begin{eqnarray}
 \label{eq:GBM}
Q_i(p)=&\beta_{\text{overall}}(p) + x_{\text{sex},i}\beta_{\text{sex}}(p) 
 +x_{\text{age},i}\beta_{\text{age}}(p) +x_{\text{DDIT3},i}\beta_{\text{DDIT3}}(p)     \notag \\
&  +x_{\text{EGFR},i}\beta_{\text{EGFR}}(p)  + x_{\text{Mesenchymal},i}\beta_{\text{Mesenchymal}}(p)  \notag \\
& 
+ x_{\text{survival},i}\beta_{ survival_{12}}(p) 
  +  E_i(p).
\end{eqnarray}

We constructed quantlets for these data using the procedure described in Section \ref{sec:quantlets}.  After the first step, we were left with a union basis set $\mathcal{D}^U$ containing 546 basis functions.   The first panel of Figure \ref{S5_Figure_qant} plots the near-losslessness parameters $\rho_0$ and $\overline{\rho}$ 
against the number of basis coefficients $K_\mathcal{C}$ in the reduced set.  Based on this, we selected the combined basis set $\mathcal{D}_\mathcal{C}$ for $\mathcal{C}=10$, which contained $K_\mathcal{C}=27$ basis functions and was near-lossless, with $\rho^0=0.990$ and $\overline{\rho}=0.998$.  We then orthogonalized, denoised, and re-standardized the resulting basis to yield the set of quantlets, the first 16 of which are plotted in Figure \ref{S5_QBE}.  As shown in panel 2 of Figure \ref{S5_Figure_qant}, these quantlets yielded a basis with similar sparsity property as principal components computed from the empirical quantile functions.

After computing the quantlet coefficients for each subject's empirical quantile function, we fit the quantlet-space version of model (\ref{eq:GBM}) as described above, obtaining $2,000$ posterior samples after a burn-in of $200$, after which the results were projected back to the original quantile space to yield posterior samples of the functional regression parameters in model (\ref{eq:GBM}).  MCMC convergence diagnostics were computed, and suggested that the chain mixed well (Supplementary Figure 17).  From these, we constructed $95\%$ point wise and joint credible bands for each $\beta_a(p)$ and computed the corresponding simultaneous band scores $P_{a,SimBaS}(p)$ and global Bayesian p-values $P_{a,Bayes}$ as described in Section \ref{sec:BayesInference}.

 \begin{figure}[!htb]
 \caption{Posterior inference for functional coefficients for T1-post contrast image:
for each covariate (6), the left panel includes posterior mean estimate, point and joint credible bands, GBPV in heading along with
SimBas less then $.05$ (orange line), 
 and the right panel includes predicted densities for the two levels of the covariate along with the posterior probability scores
   for the moment different testings.
     \label{S5_Figure_2}}
\centering
\includegraphics[height=5.3in,width=5.7in]{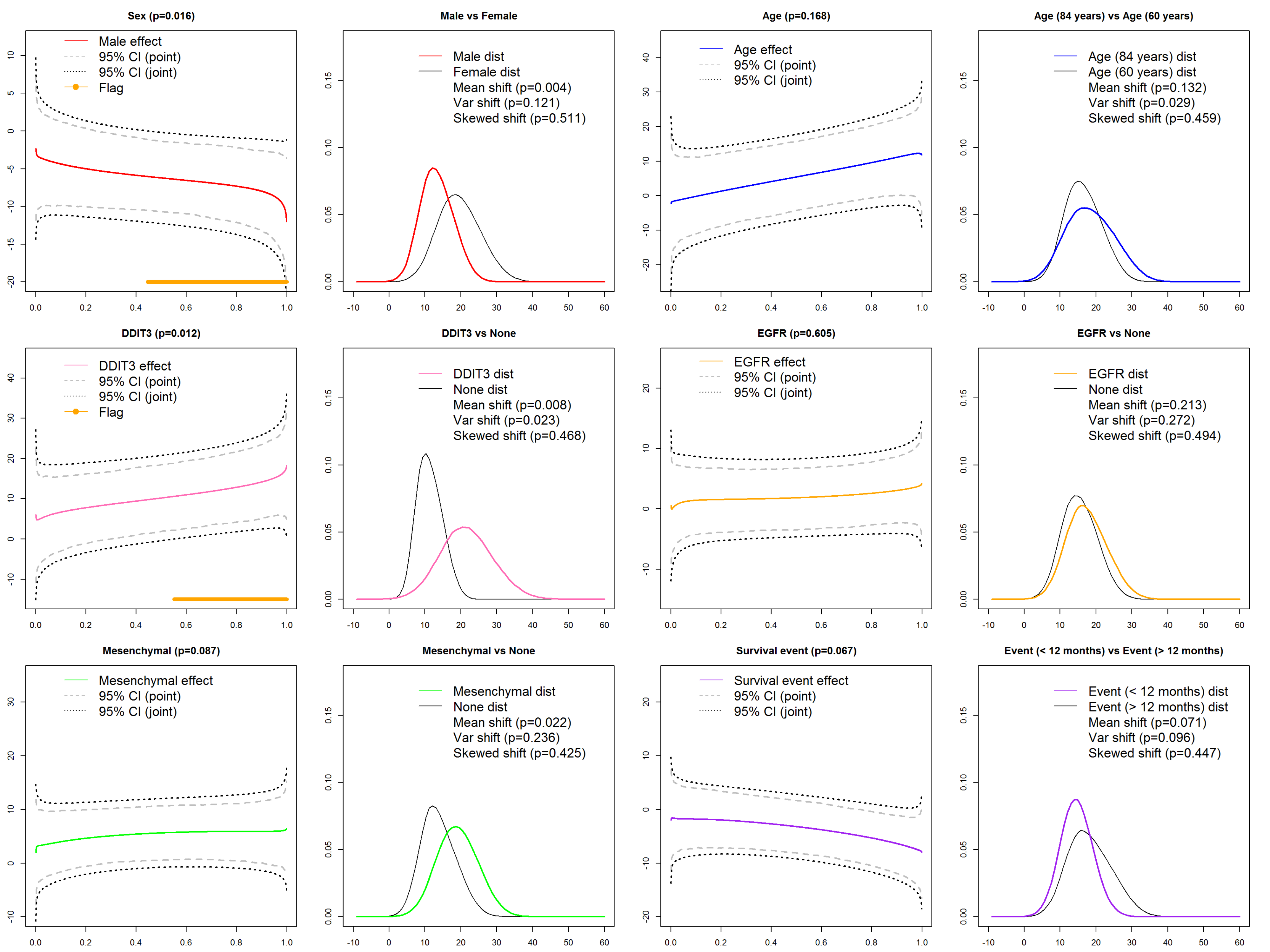}
\end{figure}

Figure  \ref{S5_Figure_2} summarizes the estimation and inference for each of the covariates in the model.  For each covariate there is one panel presenting the functional predictor $\beta_a(p)$ along with the point wise (grey) and joint (black) credible bands, and an indicator of which $p$ are flagged such that $\beta_a(p)\ne 0$ (orange lines indicating $P_{a,SimBaS}(p)<0.05$).   The other panel contains density estimates for each covariate level (holding all others at the mean), computed as outlined in the supplementary materials,  along with posterior probability scores summarizing whether the mean, variance or skewness appeared to differ across these groups.  Supplementary Table %
3
contains measures of the relative Gaussianness of the distributions for the various groups along with 95\% credible intervals.

The global Bayesian p-values for testing $\beta_a(p)\equiv0$ for each covariate are in the corresponding figure panel headers, and reveal that for sex (p=0.016) and DDIT3 (p=0.012), the functional covariates are flagged as significant, and for the mesenchymal subtype (p=0.087) and survival (p=0.067) endpoints, there was some indication of a possible trend.  We see that for sex, there was evidence of a mean shift (p=0.004) with females tending to have higher pixel intensities than males, especially in the upper tails of the distribution, and the female distribution appearing to be slightly more Gaussian than the males.  For DDIT3, we see evidence of a mean and variance shift, with tumors with DDIT3 mutation tending to have higher intensities and greater variability than those without, especially in the upper tail of the distribution.  The mesenchymal subtype, while not flagged as statistically significant in the global test, shows some tendency for a mean shift with the mesenchymal subtype tending to have higher distributional values and perhaps slightly more non-Gaussian characteristics.

Our results are presented for $K=27$ basis functions, but to assess sensitivity to choice of K we also ran our model for a wide range of possible values of $K$, with Supplementary Table 5 showing global Bayesian p-values for the entire range
of potential values for K (from 546 to 2), along with run time.  The run time tracks linearly with $K$.  Note that we get the same
substantive results over the range of basis sizes, so results are quite robust to choice of number of quantlets.  However, keeping more quantlets than necessary clearly adds to the uncertainty of parameter estimates, as indicated by the larger joint band widths. Also, keeping too few basis functions can lead to some missed results and also wider joint band widths.  Moderate basis sets that are as parsimonious as possible while retaining the near-lossless property seem to give the tightest credible bands and thus the greatest power for global and local tests.
 We also performed a sensitivity analysis on the parameter $\nu_0$ indicating the prior strength for the variance components distribution in (\ref{p5_prior_G}), and found that results for slightly larger or smaller values yielded nearly identical results. 

\begin{figure}[!htb]
 \caption{Comparison between quantlet and naive approaches for DDIT3 status for (A) quantlet approach with sparse regularization and (B) the naive \textit{one-p-at-a-time} quantile functional regression approach. 	
  \label{S5_Figure_2_g}}
\centering
 \includegraphics[height=3.1 in,width=5.7in]{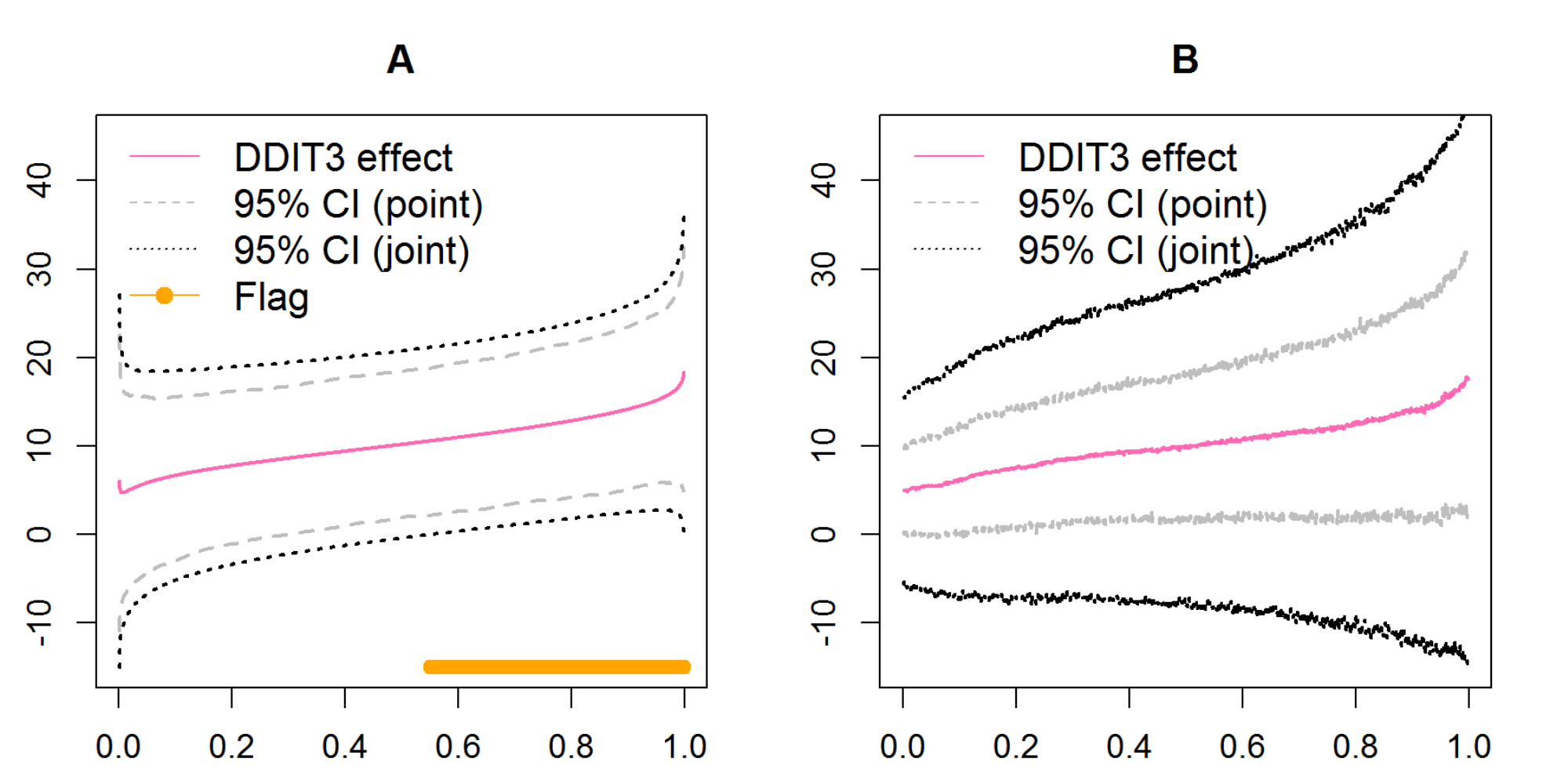}
\end{figure}

To compare different methods with our quantlet with sparse regularization approach, we also applied to these data a quantlet approach with no sparse regularization and 
a naive quantile functional regression method modeling independently for each $p$ (after interpolating onto a common grid). 
  Posterior mean estimates, credible intervals, and other inferential summaries are given in Supplementary Figure 11 and Table 6.  
    Note that the quantlets method with sparse regularization tends to yield estimates that are smoother and with tighter joint credible bands than either the naive or the quantlets-no sparse regularization runs.   As we can see in Figure~\ref{S5_Figure_2_g} the differences between the quantlet and naive methods are substantial, and demonstrate the significant power gained by borrowing strength across $p$ using the quantlet-based modeling approach.  The completely naive quantile functional regression approach gave nonsensical results for this application (Supplementary Figure 19).
    
     Supplementary Figure~18 contains the predicted quantiles functions over a grid of covariate combinations for this model.  Although the quantile functional regression using quantlets does not explicitly impose monotonocity in the predicted quantile functions, we see that the predicted quantile functions are all monotone non-decreasing. 
See Section 4 of the supplement for further details and discussion of monotonicity issues.

    Table \ref{S5_tab_pval} contains posterior probability scores assessing differences in moments for these three methods, plus a feature extraction approach in which moments were first calculated from each subject's samples and then statistically compared with a Bayesian regression fit.
As in the simulations, we see that the naive quantile functional regression method appears to have type I error problems in the mean and variance.  The quantlet results with and without sparse regularization have similar results to each other, but with the sparse regularization appearing to provide slightly smaller probability scores.  The quantlet and feature extraction approaches yielded similar results, again as in the simulations, suggesting no substantial loss of power from modeling the entire distributions if the moments are sufficient for characterizing the true differences.

\begin{table}
\caption{Posterior probability score of difference tests for the GBM data set: 
(B) naive quantile functional regression approach, 
(E) {\it quantlet} space with sparse regularization, and (G) feature extraction approach,
where the values in this table are the posterior probability scores derived by its corresponding method for each different test between treatment and reference groups in the top row.
      \label{S5_tab_pval} }
      \vspace{-0.1cm} 
\begin{center}
\resizebox{0.88\columnwidth}{!}{   
\begin{tabular}{c   |  c    c c | c c c | c c c} \hline
   $\text{{\bf Test}}$    
    &$$   &$\mu_{T}=\mu_{R}$      &$$ 
    &$$   &$\sigma_{T}=\sigma_{R}$     &$$   
    &$$    &$\xi_{T}=\xi_{R}$      &$$ 
    \\    \hline
   $\text{{\bf Method}}$     
    &$\text{{\bf B}}$   &$\text{{\bf E}}$      &$\text{{\bf G}}$ 
    &$\text{{\bf B}}$   &$\text{{\bf E}}$     &$\text{{\bf G}}$   
    &$\text{{\bf B}}$   &$\text{{\bf E}}$      &$\text{{\bf G}}$ 
    \\    \hline
 $\text{{\bf Sex}}$   &0.000     &0.004 &0.028
  &0.000     &0.121 &0.067
   &0.342     &0.511 &0.548
  \\
 $\text{{\bf Age}}$   &0.000     &0.132  &0.326
  &0.000    &0.029 &0.014
   &0.181     &0.459 &0.003
  \\
   $\text{{\bf DDIT3}}$   &0.000    &0.008  &0.020
  &0.000    &0.023 &0.046
   &0.347   &0.468 &0.442
  \\
   $\text{{\bf EGFR}}$   &0.000    &0.213  &0.470
  &0.000    &0.272 &0.391
   &0.365     &0.494 &0.470
  \\
   $\text{{\bf Mesenchymal}}$   &0.000     &0.022 &0.043
  &0.000    &0.236 &0.458
   &0.071     &0.425 &0.189
  \\
     ${\bf Survival_{12}}$   &0.000   &0.071 &0.160
  &0.000    &0.096 &0.034
   &0.309    &0.447 &0.941
  \\  \hline
\end{tabular}
}
\end{center}
\end{table}      


\section{{\bf Discussion }}

In this paper, we have introduced a strategy for regressing the distribution of repeated samples for a subject on a set of covariates through a model we call \textit{quantile functional regression}.  We distinguish this model from other types of quantile regression and functional regression  methods in existing literature, in that it is regressing the \textit{subject-specific} quantile, not the \textit{population-level} quantile, on covariates, and accounts for intrasubject correlation.  We describe how it serves as a middle ground between two commonly-used strategies of (1) performing a series of regressions on arbitrary summaries of the distribution such as mean or standard deviation and (2) independent regression models for each quantile $p$ in a chosen set.  Our approach models a subject's entire quantile function as a functional response, building in dependency across $p$ in the mean and covariance using custom basis functions called \textit{quantlets} that are empirically defined, near-lossless, regularized, sparse, and with some of the individual bases being interpretable.  These basis functions have sparsity properties similar to principal components, but appear more regular and interpretable.  They provide a flexible representation of the underlying quantile functions while containing a sufficient Gaussian basis as a subspace.

We show by simulation and in our GBM application that the use of the quantlets has significant advantages over naive \textit{one-p-at-a-time} quantile functional regression approaches through smoother estimates and greater power via tighter joint credible bands and better type I error properties when comparing distributional moments.  Yet, it also appears to experience essentially no power loss to detect location and scale shifts relative to feature extraction approaches when distributional differences can be thus characterized.    In this manner, our approach can detect many types of differences in the underlying distributions, even those not captured by differences in the moments, without sacrificing power when the differences are straightforward enough to be summarized by the moments.

We fit the quantile functional  regression model using a Bayesian approach with sparsity regularization priors on the quantlet space regression coefficients that smooths the regression coefficients and yields a broad array of Bayesian inferential summaries computable from the posterior samples of the MCMC procedure.   
For example, we can construct global tests of signficance for each covariates using global Bayesian p-values, and then characterize these difference by flagging regions of $p$ while adjusting for multiple testing, and obtaining probability scores for any moments or other summaries of the distributions.

While we use Bayesian model with sparse regularization via spike-slab prior for the reasons given above, the general strategy of performing quantile functional regression using quantlets could also be paired with other functional regression approaches, including frequentist approaches that use penalized likelihood fitting for regularization, as our only informative priors were the regularization parameters that are estimated from the data.

In the supplement, we provide R scripts to compute quantlets and fit the quantile functional regression model.  We also share the GBM application data and R scripts to run the analyses and produce the plots contained within this paper.

In this paper, we have presented the quantile functional regression framework using a standard linear model with scalar covariates and independent Gaussian residual error functions, but as in other functional regression contexts the model can be extended to include other complex structures that extend the usability of the modeling framework.  This includes functional covariates, nonparametric effects in the covariates $x_{ia}$, random effects and/or spatially/temporally correlated residual errors to accommodate correlation between subjects induced by the experimental design, and the ability to perform robust quantile functional regression to downweight outlying samples using heavier-tailed likelihoods. These types of flexible modeling components are available as part of the Bayesian functional mixed model (BayesFMM) methods that have been developed in recent years
\cite{morris2006wavelet,zhu2011robust,zhu2012robust,meyer2015bayesian,zhang2016functional,Zhu2017robustf,lee2016semiparametric}.
By linking the software developed here to generate the quantlets and fit quantile functional regression models with the BayesFMM software, it will be possible to extend the quantile functional regression framework to these settings and thus analyze an even broader array of complex data sets generated by modern research tools.

Our approach has been designed with relatively high dimensional data in mind, i.e. data for which there are at least 
a moderately large number of observations per subject (at least 50 or 100). 
  We are currently working on extensions of this method to handle lower dimensional data with fewer observations per subject, which requires a careful propagation of uncertainty in the estimators of the empirical quantile functions into the quantile functional regression.  This propagation of uncertainty could also be done in larger sample cases like the one presented here, but given the substantial complexity and length already in this paper we leave this for future work.  
Also, in settings with enormous numbers of observations per subjects, e.g. millions to billions or more, the procedure described in this paper to construct the quantlets basis would be too computationally burdensome.  Given that in those settings, it is unlikely that so many observations are needed to quantify the subject-specific quantile function, we have worked out algorithms to down-sample the empirical quantile functions in these cases in a way that engenders computational feasibility but is still near-lossless.  This also will be reported in future work.   Other data have measurements on many 1000s to 100,000s of subjects, which can be accommodated by computational adjustments of the procedure reported herein, but again we leave this for future work.   In this paper, we focused on absolutely continuous random variables that have no jumps in the quantile functions.  It is also possible to adapt our quantlet construction procedure to allow jumps at a discrete set of values, thus accommodating discrete valued random variables, but again this extension will be left for future work.

We believe that as more applications yield large number of automated measurements for each individual, the quantile functional regression framework we introduce in this paper may have a great impact on many areas of science to extract information and discover relationships in big, complex data.

\baselineskip 12pt
\bibliographystyle{plain}
\bibliography{ms1.bib}

\begin{thebibliography}{10}

\bibitem{armagan2013generalized}
Artin Armagan, David~B Dunson, and Jaeyong Lee.
\newblock Generalized double pareto shrinkage.
\newblock {\em Statistica Sinica}, 23:119--143, 2013.

\bibitem{bhattacharya2015dirichlet}
Anirban Bhattacharya, Debdeep Pati, Natesh~S Pillai, and David~B Dunson.
\newblock Dirichlet--laplace priors for optimal shrinkage.
\newblock {\em Journal of the American Statistical Association},
  110:1479--1490, 2015.

\bibitem{brockhaus2015fdboost}
S~Brockhaus and D~R{\"u}gamer.
\newblock Fdboost: boosting functional regression models.
\newblock {\em R package version 0.0-8}, 2015.

\bibitem{brockhaus2015functional}
Sarah Brockhaus, Fabian Scheipl, Torsten Hothorn, and Sonja Greven.
\newblock The functional linear array model.
\newblock {\em Statistical Modelling}, 15:279--300, 2015.

\bibitem{cardot2005quantile}
Herv{\'e} Cardot, Christophe Crambes, and Pascal Sarda.
\newblock Quantile regression when the covariates are functions.
\newblock {\em Nonparametric Statistics}, 17:841--856, 2005.

\bibitem{carvalho2010horseshoe}
Carlos~M Carvalho, Nicholas~G Polson, and James~G Scott.
\newblock The horseshoe estimator for sparse signals.
\newblock {\em Biometrika}, 97:465--480, 2010.

\bibitem{chen2012conditional}
Kehui Chen and Hans-Georg M{\"u}ller.
\newblock Conditional quantile analysis when covariates are functions, with
  application to growth data.
\newblock {\em Journal of the Royal Statistical Society: Series B}, 74:67--89,
  2012.

\bibitem{davino2013quantile}
Cristina Davino, Marilena Furno, and Domenico Vistocco.
\newblock {\em Quantile regression: theory and applications}.
\newblock Wiley New York, 2013.

\bibitem{donoho1995wavelet}
David~L Donoho, Iain~M Johnstone, G{\'e}rard Kerkyacharian, and Dominique
  Picard.
\newblock Wavelet shrinkage: asymptopia?
\newblock {\em Journal of the Royal Statistical Society. Series B},
  57:301--369, 1995.

\bibitem{dunson2006bayesian}
David~B Dunson.
\newblock Bayesian dynamic modeling of latent trait distributions.
\newblock {\em Biostatistics}, 7:551--568, 2006.

\bibitem{dunson2007bayesian}
David~B Dunson, Natesh Pillai, and Ju-Hyun Park.
\newblock Bayesian density regression.
\newblock {\em Journal of the Royal Statistical Society: Series B},
  69:163--183, 2007.

\bibitem{faraway1997regression}
Julian~J Faraway.
\newblock Regression analysis for a functional response.
\newblock {\em Technometrics}, 39:254--261, 1997.

\bibitem{felipe2013cancer}
E~Melo Felipe De~Sousa, Louis Vermeulen, Evelyn Fessler, and Jan~Paul Medema.
\newblock Cancer heterogeneity - a multifaceted view.
\newblock {\em EMBO reports}, 14:686--695, 2013.

\bibitem{ferraty2005conditional}
Fr{\'e}d{\'e}ric Ferraty, Abbes Rabhi, and Philippe Vieu.
\newblock Conditional quantiles for dependent functional data with application
  to the climatic" el ni{\~n}o" phenomenon.
\newblock {\em Sankhy{\=a}: The Indian Journal of Statistics}, 67:378--398,
  2005.

\bibitem{goldsmith2011penalized}
Jeff Goldsmith, Jennifer Bobb, Ciprian~M Crainiceanu, Brian Caffo, and Daniel
  Reich.
\newblock Penalized functional regression.
\newblock {\em Journal of Computational and Graphical Statistics}, 20:830--851,
  2012.

\bibitem{goldsmith2011functional}
Jeff Goldsmith, Matt~P Wand, and Ciprian Crainiceanu.
\newblock Functional regression via variational bayes.
\newblock {\em Electronic journal of statistics}, 5:507--602, 2011.

\bibitem{griffin2010inference}
Jim~E Griffin, Philip~J Brown, et~al.
\newblock Inference with normal-gamma prior distributions in regression
  problems.
\newblock {\em Bayesian Analysis}, 5:171--188, 2010.

\bibitem{griffin2006order}
Jim~E Griffin and MF~J Steel.
\newblock Order-based dependent dirichlet processes.
\newblock {\em Journal of the American statistical Association}, 101:179--194,
  2006.

\bibitem{guo2002functional}
Wensheng Guo.
\newblock Functional mixed effects models.
\newblock {\em Biometrics}, 58:121--128, 2002.

\bibitem{hao2007quantile}
Lingxin Hao and Daniel~Q Naiman.
\newblock {\em Quantile regression}.
\newblock Sage London, 2007.

\bibitem{he2000quantile}
Xuming He and Hua Liang.
\newblock Quantile regression estimates for a class of linear and partially
  linear errors-in-variables models.
\newblock {\em Statistica Sinica}, 10:129--140, 2000.

\bibitem{just2014improving}
Nathalie Just.
\newblock Improving tumour heterogeneity mri assessment with histograms.
\newblock {\em British journal of cancer}, 111:2205--2213, 2014.

\bibitem{kato2012estimation}
Kengo Kato.
\newblock Estimation in functional linear quantile regression.
\newblock {\em The Annals of Statistics}, 6:3108--3136, 2012.

\bibitem{kato2012asymptotics}
Kengo Kato, Antonio~F Galvao, and Gabriel~V Montes-Rojas.
\newblock Asymptotics for panel quantile regression models with individual
  effects.
\newblock {\em Journal of Econometrics}, 170:76--91, 2012.

\bibitem{koenker2004quantile}
Roger Koenker.
\newblock Quantile regression for longitudinal data.
\newblock {\em Journal of Multivariate Analysis}, 91:74--89, 2004.

\bibitem{koenker2005quantile}
Roger Koenker.
\newblock {\em Quantile regression}.
\newblock Cambridge university press, 2005.

\bibitem{lawrence1989concordance}
I~Lawrence and Kuei Lin.
\newblock A concordance correlation coefficient to evaluate reproducibility.
\newblock {\em Biometrics}, 45:255--268, 1989.

\bibitem{lee2016semiparametric}
Wonyul Lee, Michelle Miranda, Veera Baladandayuthapani, P~Rausch, M~Fazio,
  C~Downs, and Jeffrey~S Morris.
\newblock Semiparametric functional mixed models for longitudinal functional
  data, with application to glaucoma data.
\newblock {\em Under revision}, 2017.

\bibitem{lempers1971posterior}
Fred~B Lempers.
\newblock {\em Posterior probabilities of alternative linear models}.
\newblock Rotterdam University Press, 1971.

\bibitem{li2016inference}
Meng Li, Kehui Wang, Arnab Maity, and Ana-Maria Staicu.
\newblock Inference in functional linear quantile regression.
\newblock {\em arXiv preprint arXiv:1602.08793}, 2016.

\bibitem{maceachern1999dependent}
Steven~N MacEachern.
\newblock Dependent nonparametric processes.
\newblock {\em ASA proceedings of the section on Bayesian statistical science},
  1:50--55, 1999.

\bibitem{marusyk2012intra}
Andriy Marusyk, Vanessa Almendro, and Kornelia Polyak.
\newblock Intra-tumour heterogeneity: a looking glass for cancer?
\newblock {\em Nature Reviews Cancer}, 12:323--334, 2012.

\bibitem{meyer2015bayesian}
Mark~J Meyer, Brent~A Coull, Francesco Versace, Paul Cinciripini, and Jeffrey~S
  Morris.
\newblock Bayesian function-on-function regression for multilevel functional
  data.
\newblock {\em Biometrics}, 71:563--574, 2015.

\bibitem{mitchell1988bayesian}
Toby~J Mitchell and John~J Beauchamp.
\newblock Bayesian variable selection in linear regression.
\newblock {\em Journal of the American Statistical Association}, 83:1023--1032,
  1988.

\bibitem{morris2015functionalreg}
Jeffrey~S Morris.
\newblock Functional regression.
\newblock {\em Annual Review of Statistics and Its Application}, 2:321--359,
  2015.

\bibitem{morris2006wavelet}
Jeffrey~S Morris and Raymond~J Carroll.
\newblock Wavelet-based functional mixed models.
\newblock {\em Journal of the Royal Statistical Society. Series B},
  68:179--199, 2006.

\bibitem{muller1996bayesian}
Peter Muller, Alaattin Erkanli, and Mike West.
\newblock Bayesian curve fitting using multivariate normal mixtures.
\newblock {\em Biometrika}, 83:67--79, 1996.

\bibitem{park2008bayesian}
Trevor Park and George Casella.
\newblock The bayesian lasso.
\newblock {\em Journal of the American Statistical Association}, 103:681--686,
  2008.

\bibitem{parzen2004quantile}
Emanuel Parzen.
\newblock Quantile probability and statistical data modeling.
\newblock {\em Statistical Science}, 19:652--662, 2004.

\bibitem{ramsay2006functional}
James~O Ramsay and B.~W. Silverman.
\newblock {\em Functional data analysis}.
\newblock New York: Springer, 2006.

\bibitem{reich2012spatiotemporal}
Brian~J Reich.
\newblock Spatiotemporal quantile regression for detecting distributional
  changes in environmental processes.
\newblock {\em Journal of the Royal Statistical Society: Series C},
  61:535--553, 2012.

\bibitem{reich2012bayesian}
Brian~J Reich, Montserrat Fuentes, and David~B Dunson.
\newblock Bayesian spatial quantile regression.
\newblock {\em Journal of the American Statistical Association}, 106:6--20,
  2012.

\bibitem{reiss2010fast}
Philip~T Reiss, Lei Huang, and Maarten Mennes.
\newblock Fast function-on-scalar regression with penalized basis expansions.
\newblock {\em International Journal of Biostatistics}, 6:1--28, 2010.

\bibitem{ruppert2003semiparametric}
David Ruppert, Matt~P Wand, and Raymond~J Carroll.
\newblock {\em Semiparametric regression}.
\newblock Cambridge university press, 2003.

\bibitem{saha2016demarcate}
Abhijoy Saha, Sayantan Banerjee, Sebastian Kurtek, Shivali Narang, Joonsang
  Lee, Ganesh Rao, Juan Martinez, Karthik Bharath, Arvind~UK Rao, and
  Veerabhadran Baladandayuthapani.
\newblock Demarcate: Density-based magnetic resonance image clustering for
  assessing tumor heterogeneity in cancer.
\newblock {\em NeuroImage: Clinical}, 12:132--143, 2016.

\bibitem{scheipl2015functional}
Fabian Scheipl, Ana-Maria Staicu, and Sonja Greven.
\newblock Functional additive mixed models.
\newblock {\em Journal of Computational and Graphical Statistics}, 24:477--501,
  2015.

\bibitem{tibshirani1996regression}
Robert Tibshirani.
\newblock Regression shrinkage and selection via the lasso.
\newblock {\em Journal of the Royal Statistical Society. Series B},
  58:267--288, 1996.

\bibitem{tutt2011glioblastoma}
B~Tutt.
\newblock Glioblastoma cure remains elusive despite treatment advances.
\newblock {\em OncoLog}, 56:1--8, 2011.

\bibitem{wu2000kernel}
Colin~O Wu and Chin-Tsang Chiang.
\newblock Kernel smoothing on varying coefficient models with longitudinal
  dependent variable.
\newblock {\em Statistica Sinica}, 10:433--456, 2000.

\bibitem{yang2017joint}
Yun Yang and Surya~T Tokdar.
\newblock Joint estimation of quantile planes over arbitrary predictor spaces.
\newblock {\em Journal of the American Statistical Association}, pages 1--14,
  2017.

\bibitem{yang2015quantile}
Yunwen Yang and Xuming He.
\newblock Quantile regression for spatially correlated data: an empirical
  likelihood approach.
\newblock {\em Statistica Sinica}, 25:261--274, 2015.

\bibitem{zhang2016functional}
Lin Zhang, Veerabhadran Baladandayuthapani, Hongxiao Zhu, Keith~A Baggerly,
  Tadeusz Majewski, Bogdan~A Czerniak, and Jeffrey~S Morris.
\newblock Functional car models for large spatially correlated functional
  datasets.
\newblock {\em Journal of the American Statistical Association}, 111:772--786,
  2016.

\bibitem{Zhu2017robustf}
H~Zhu, F~Versace, P~Cinciripini, and Jeffrey~S Morris.
\newblock Robust functional mixed models for spatially correlated functional
  regression, with application to event-related potentials for
  nicotine-addicted individuals.
\newblock {\em Under review}, 2017.

\bibitem{zhu2011robust}
Hongxiao Zhu, Philip~J Brown, and Jeffrey~S Morris.
\newblock Robust, adaptive functional regression in functional mixed model
  framework.
\newblock {\em Journal of the American Statistical Association},
  106:1167--1179, 2011.

\bibitem{zhu2012robust}
Hongxiao Zhu, Philip~J Brown, and Jeffrey~S Morris.
\newblock Robust classification of functional and quantitative image data using
  functional mixed models.
\newblock {\em Biometrics}, 68:1260--1268, 2012.

\end{thebibliography}

\end{document}